\documentclass[preprint,superscriptaddress,frontmatterverbose,aps,prd,amsmath,amssymb,floatfix,notitlepage,nofootinbib,longbibliography]{revtex4-2}

\usepackage{xcolor}
\usepackage{graphicx}
\usepackage{dcolumn}
\usepackage{orcidlink}
\usepackage{cancel}

\allowdisplaybreaks

\def \hc{\mathrm{h.c.}}
\def \ts{\hskip 0.08em}
\def \tts{\hskip 0.04em}

\def \nn{\nonumber}

\def \cB{{\cal B}}

\def \hc{\,+~{\rm h.c.}}
\def \eff{{\rm eff}}

\def\beq{\begin{equation}}
\def\eeq{\end{equation}}
\def\bea{\begin{eqnarray}}
\def\eea{\end{eqnarray}}
\def\nn{\nonumber}
\def\roughly#1{\mathrel{\raise.3ex\hbox
{$#1$\kern-.75em\lower1ex\hbox{$\sim$}}}}

\def\gsim{\roughly>}

\def\bctaunu{b \to c \tau^- {\bar\nu}_\tau}
\def\order{\lower 1.8ex \hbox{\LARGE\~{}}}
\def\btopiK{B \to  K \pi}

\begin{document}
\count\footins = 1000

\title{Flavor Violations in $B$-Mesons within Non-Minimal SU(5)}
\author{Bhubanjyoti Bhattacharya\,\orcidlink{0000-0003-2238-321X}}
\email{bbhattach@ltu.edu}
\affiliation{Department of Natural Sciences, Lawrence Technological University, Southfield, MI 48075, USA}

\author{Alakabha Datta\,\orcidlink{0000-0001-8713-2783}}
\email{datta@phy.olemiss.edu}
\affiliation{Department of Physics and Astronomy,
108 Lewis Hall, University of Mississippi, Oxford, MS 38677-1848, USA}

\author{Gaber Faisel\,\orcidlink{0000-0001-8770-1966}}
\email{gaberfaisel@sdu.edu.tr}
\affiliation{Department of Physics, Faculty of Engineering and Natural Sciences,
S\"uleyman Demirel University, Isparta, Turkey 32260}
\altaffiliation{deceased}

\author{Shaaban Khalil\,\orcidlink{0000-0003-1950-4674}}
\email{skhalil@zewailcity.edu.eg}
\affiliation{Center for Fundamental Physics, Zewail City of Science and
Technology, 6th of October City, Giza 12578, Egypt}

\author{Shibasis Roy\,\orcidlink{0000-0002-0744-9113}}
\email{shibasis.cmi@gmail.com}
\affiliation{Chennai Mathematical Institute, Siruseri 603103, Tamil Nadu, India}

\begin{abstract}
Recent anomalies in $B$-meson decays, such as deviations in $R_{D^{(*)}}$ and $B\to K\nu{\bar\nu}$, suggest possible lepton flavor universality violation and new exotic interactions. In this work, we explore these anomalies within a non-minimal SU(5) grand unified theory (GUT) framework, which introduces a 45-dimensional Higgs representation predicting exotic scalar particles, including the leptoquark $R_2$ and diquark $S_6$. The $R_2$ leptoquark addresses charged current anomalies in $b\to c\tau\nu$ transitions, the $S_6$ diquark contributes to nonleptonic neutral current processes, such as $B\to K\pi$, while at the loop level, diagrams involving the exchanges of the leptoquark, diquark, and the Standard Model particles contribute to $B\to K\nu{\bar\nu}$, offering solutions to longstanding puzzles.
\end{abstract}

\maketitle
\newpage

\section{Introduction}

\setlength{\footnotesep}{2em}

The masses and mixing of the quarks and leptons, along with the unknown reason for three generations, are the pressing flavor puzzles in the Standard Model (SM) that remain unresolved. In addition, the gauge interactions of the three families of quarks and leptons are the same. This forbids tree-level flavor-changing neutral current (FCNC) processes. At the loop level, FCNC effects in $b$ quark decays are enhanced due to contributions from the heavy top quark. However, these processes are still rare in the SM, allowing for new physics effects to be present. Hence, understanding FCNC effects in $B$-meson decays provides crucial insights into the physics beyond the SM. Among the various theoretical extensions of the SM, the SU(5) grand unified theory (GUT) stands out as one of the most compelling. It unifies the SM gauge groups into a single framework and, combines quarks and leptons into a single multiplet, thereby predicting electric charge quantization. SU(5) provides a promising avenue for understanding fundamental interactions by linking the strong, weak, and electromagnetic forces. However, minimal SU(5) faces several significant challenges. For instance, it does not successfully unify the SM gauge couplings and predicts incorrect fermion mass relations that conflict with experimental data. A potential solution to address some of these issues is to introduce an additional Higgs multiplet with a 45-dimensional representation \cite{Georgi:1979df, Babu:1984vx}. This model also enables a richer flavor structure and introduces potential new sources of flavor violation. In particular, the inclusion of exotic scalars, such as the leptoquark \( R_2 \) and diquark \( S_6 \), significantly broadens the scope of potential flavor-violating interactions in non-minimal SU(5) models. These particles, which emerge naturally in the extended SU(5) framework, introduce interactions beyond those allowed in the SM, offering a pathway to address key experimental anomalies and deepen our understanding of flavor physics.

Studies of $B$-meson decays have played an important role in establishing the flavor structure of the SM. FCNC processes in $b$-quark decays are also crucial in the search for new physics, as they are highly sensitive to new sources of CP violation and exotic particles. Over several decades, many $B$-decay measurements have shown deviations from their SM predictions -- these are known as the $B$ anomalies. The anomalies can be classified as charged-current and neutral-current anomalies. In the charged-current sector, the measurements of interest are the ratios $R_{D^{(*)}} \equiv \cB({\bar B}\to D^{(*)}\tau^-{\bar\nu_\tau})/\cB({\bar B}\to D^{(*)}\ell^-{\bar\nu_\ell})$ (here $\ell = e,\mu$) and $R_{J/\psi} \equiv \cB(B_c^+\to J/\psi\tau^+\nu_\tau)/\cB(B_c^+\to J/\psi\mu^+\nu_\mu)$. The first of these ratios, $R_{D^{(*)}}$, has been measured with a precision of 5-8\% \cite{Lees:2012xj, Lees:2013uzd, Aaij:2015yra, Huschle:2015rga, Sato:2016svk, Hirose:2016wfn, Aaij:2017uff, Hirose:2017dxl, Aaij:2017deq, Belle:2019gij, LHCb:2023zxo, LHCb:2023uiv, Belle-II:2024ami, LHCb:2024jll}, by the BaBar, Belle, and LHCb experiments. These measurements show a combined deviation of 3.31$\sigma$ \cite{HFLAV:RDRDst2024update} from the SM\footnote{Recent experimental updates have increased the combined deviation from the SM in $R_D$ and $R_{D^*}$ to $\sim$ 3.8$\sigma$ \cite{HFLAV:2024ctg}. While we have chosen to use observable values from the 2024 HFLAV update for our numerical results, the larger deviations will only make the case for new physics stronger without affecting the essential physics arguments made in this paper.}. The second ratio, $R_{J/\psi}$, has been measured by the LHCb experiment with a precision of 35\% \cite{Aaij:2017tyk} and displays a 1.7$\sigma$ deviation from its SM prediction\footnote{The CMS collaboration has also measured $R_{J/\psi}$ in leptonic \cite{CMS:2024seh} and hadronic \cite{CMS:2025jfx} $\tau$-decay channels. While these measurements are consistent with the SM prediction, the precision of these measurements is low ($\gtrsim$ 50\%) and are not used as inputs in this paper.}. Together, these measurements provide strong hints of lepton-universality-violating new physics in charged-current $b$ decays.

In the neutral-current sector, there are anomalies in semileptonic and nonleptonic $B$ decays. In the semileptonic $B$ decays, there are several hints of deviations from the SM in the FCNC $b\to s$ transitions with charged leptons and neutrinos in the final state. A recent first measurement of the branching ratio $\cB(B^+\to K^+\nu\bar\nu) = (2.3\pm 0.7)\times 10^{-5}$ by the Belle II experiment~\cite{Belle-II:2023esi} is $2.7\sigma$ higher than the SM expectation
$\cB(B^+\to K^+ \nu\bar\nu)_{\rm SM}=(5.58\pm 0.38)\times 10^{-6}$~\cite{Parrott:2022zte}. Furthermore, even though the recently updated measurements of $R_{K^{(*)}} = \cB(B\to K^{(*)} \mu^+\mu^-)/\cB(B\to K^{(*)} e^+ e^-)$ are now fully consistent with their SM expectations~\cite{LHCb:2022qnv}, the individual branching fractions in both the electron and muon channels remain discrepant~\cite{Capdevila:2023yhq}). 
There has also been recent interest in several anomalies in neutral-current hadronic $B$ decays \cite{Beaudry:2017gtw,Kundu:2021emt,Bhattacharya:2021shk,Bhattacharya:2022akr,Amhis:2022hpm,Biswas:2023pyw,Berthiaume:2023kmp,Altmannshofer:2024kxb,Grossman:2024amc}. In particular, we will focus on the hadronic $B$-decay anomaly that has been around for about 20
years, known as the $\btopiK$ puzzle (see Refs.~\cite{Buras:2003yc,Baek:2004rp,Beaudry:2017gtw,  Bhattacharya:2021shk} and references therein). Here the amplitudes for the four decays $B^+ \to \pi^+ K^0$, $B^+ \to \pi^0 K^+$, $B^0 \to
\pi^- K^+$ and $B^0 \to \pi^0 K^0$ obey a quadrilateral isospin relation. However, the measurements of the observables in these decays are not completely consistent with one another -- there is a discrepancy at the level of $\sim 3\sigma$.

The non-minimal SU(5) model proposed in this work has several new states that can potentially resolve the various $B$ anomalies. The leptoquark \( R_2 \), with its ability to mediate interactions between quarks and leptons, can explain observed anomalies in semileptonic decays, particularly the $R^{\tau/\ell}_{D^{(*)}}$ ratios in $b\to c\tau\nu$ transitions. The \( R_2 \) leptoquark, with its distinct coupling patterns, presents a viable candidate for reconciling these discrepancies and, if confirmed, could establish a direct link between grand unification and lepton-flavor universality violation. On the other hand, the diquark \( S_6 \) enables novel quark-quark interactions that are absent in the SM. This scalar diquark has the potential to contribute significantly to flavor-changing neutral current processes in $B$-meson decays, such as \( B \to K \pi \). The leptoquark and diquark can appear at the loop level with the SM particles and contribute to \( B \to K \nu \bar{\nu} \), thereby producing  new physics signals in these decays. A model with diquarks and leptoquarks that resolves the $B$ anomalies was discussed in Ref.~\cite{Datta:2019tuj}. 

In this work, we investigate the role of exotic scalar fields in flavor-violating $B$-meson decays within the non-minimal SU(5) model. We focus on the contributions of the diquark \( S_6 \) and leptoquark \( R_2 \) to rare decay channels, aiming to shed light on their potential to address current experimental anomalies. By analyzing the \( B \to K \pi \), \( b \to c \tau \nu \), and \( B \to K \nu \nu \) processes, we provide insights into how these exotic scalars may offer a promising avenue for discovering new physics through flavor violation.

The paper is structured as follows. In Section 2, we briefly introduce minimal SU(5), focusing on the main features of the light exotic 45-dimensional scalars, specifically the leptoquark \( R_2 \) and diquark \( S_6 \). In Section 3, we examine the contributions of the diquark \( S_6 \) to the \(B\to K\pi\) processes and determine the best-fit parameters consistent with experimental measurements. Section 4 addresses the contribution of the leptoquark \( R_2 \) to the \( b \to c \tau \nu \) process, highlighting its potential to account for the \( R_D \) and \( R_{D^*} \) anomalies. In Section 5, we consider the combined contributions of both \( S_6 \) and \( R_2 \) to \( B \to K \nu \nu \), a process with a very small SM prediction, making it a promising channel for probing new physics through flavor violation. Finally, we present our conclusions and future prospects in Section 6. In the appendix, we address the issue of gauge coupling unification in our model.

\section{Non-Minimal SU(5)}
\label{sec:nonminSU5}

In this section, we describe the Non-Minimal SU(5) model, which includes a Higgs in the $45$ representation. Starting from the Lagrangian for the model, we discuss constraints on the model parameters from FCNC $b\to s(d) \gamma$ processes.

As advocated, extending the Higgs sector of SU(5) by $45_H$ helps solve some of the problems that the minimal SU(5) GUT faces \cite{Dorsner:2006dj,FileviezPerez:2007nh, Khalil:2014gba,Ismael:2023ffx}. The $45_H$ transforms under the SM gauge group, SU(3)$_{\rm C}\times$SU(2)$_{\rm L}\times$U(1)$_Y$, as follows. 
\beq %
45_H = (8,2)_{1/2}\oplus (1,2)_{1/2}\oplus (3,1)_{-1/3} \oplus
(3,3)_{-1/3}
 \oplus  (6^*,1){-1/3}\oplus (3^*,2)_{-7/6}\oplus
(3^*,1)_{4/3}.
\eeq%
It also satisfies the following constraints: $45^{\alpha
\beta}_\gamma = - 45^{\beta \alpha}_\gamma$ and $\sum_\alpha^5
(45)^{\alpha \beta}_\alpha =0$. Through non-vanishing Vacuum Expectation Values (VEVs) of $5_H$ and $45_H$: $\langle 5_H
\rangle = v_5,\langle 45_H \rangle^{15}_1 = \langle 45_H
\rangle^{25}_2 = \langle 45_H \rangle^{35}_3 = v_{45}, \langle
45_H \rangle^{45}_4 = -3 v_{45}$, the electroweak symmetry
$SU(2)_L \times U(1)_Y$ is spontaneously broken into $U(1)_{\rm EM}$.
We present the decomposition of the $45_H$ scalar field in Table \ref{Spec}. 
\begin{table}[ht]
\centering
\begin{tabular}{|c|c|c|}
\hline
\hline
 $(\bar{6},1,-1/3)^{ij}_k$~ &~ $S^*_6 $ ~&~ $\phi^{ij}_k \equiv \frac{1}{6}~ \epsilon^{ijl} \phi_{lk}$   \\ [1ex]
\hline
$(\bar{3},2,-7/6)^{ij}_c$ & $R_2^*$ & $ \phi^{ij}_c\equiv \frac{1}{6}~ \epsilon^{ijl} \phi_{lc}=\frac{1}{6}~ \epsilon^{ijl} (45_H)_{lc}$    \\[1ex]
\hline
$(\bar{3},1,4/3)^{ab}_k$ & $\tilde{S}_1 $ & $\phi^{ab}_k \equiv   \frac{1}{2}~ \delta^{ab} (45_H)_{k}$    \\[1ex]
\hline
$(3,1,-1/3)^{ib}_c$ & $S_1$ & $\phi^{ib}_c\equiv \frac{1}{2}~ \delta^b_c (45_H)^i =- \frac{1}{2}~ \delta^j_k (45_H)^i$ \\
[0.5ex]
$(3,1,1/3)^{ij}_k$  &  &   \\[1ex]
\hline
$(3,3,-1/3)^{ib}_c$ & $S_3$  & $\phi^{ib}_c\equiv (45_H)^{ib}_c -\frac{1}{2} \delta^b_c (45_H)^{id}_d $  \\[1ex]
\hline
$(1,2,1/2)^{ab}_c$ &  $D_2$ &  $\phi^{ab}_c(\phi^{ib}_k)\equiv \frac{1}{2} \delta^a_c (45_H)^b \equiv -\frac{1}{2}~ \delta^i_k (45_H)^b$   \\[0.5ex]
$(1,2,1/2)^{ib}_k$ & & \\[1ex]
\hline
$ (8,2,1/2)^{ia}_j$ & $S_8$ & $\phi^{ia}_j \equiv (45_H)^{ia}_j -\frac{1}{3} \delta^i_j (45_H)^{ka}_k $  \\[1ex]
\hline
\hline
\end{tabular}
\caption{ \label{Spec} 45-Higgs scalar spectrum, with \(SU(3)\) indices \(i, j, k\) and \(SU(2)\) indices \(a, b, c\). }
\end{table}

In general, the Yukawa interactions of the $45_H$ field can be expressed as \cite{Goto:2023qch}:
\beq
{-\mathcal{L}_Y
  \!=\! \frac{1}{4}(Y_{45}^U)_{ij}\epsilon_{ABCDE}\ts
    (\Psi_{10\ts i})^{AB}(\Phi_{45})^{CD}_F(\Psi_{10\ts j})^{EF}
  \!+\! \frac{1}{2}\ts (Y_{45}^D)_{ij}
    (\Psi_{10\ts i})^{AB}(\Phi_{45}^\dagger)_{AB}^C(\Psi_{\bar{5}\tts j})_C
  \!+\! \mathrm{h.c.}\,,}
  \label{eq:SU5Yukawa}
\eeq%
where $\epsilon_{ABCDE}$ is the totally antisymmetric tensor with $\epsilon_{12345} = 1$, and $Y_{45}^U$ is an antisymmetric matrix in generation space that satisfies
\begin{align}
(Y_{45}^U)_{ij}=-(Y_{45}^U)_{ji}\,,
\end{align}
where $i, j$ are generation indices.

After SU(5) symmetry breaking, the Lagrangian, involving the $R_2$ and $S_6$ fields, in Eq.~(\ref{eq:SU5Yukawa}) can be expanded as,
\cite{Goto:2023qch}
\bea
- \mathcal{L}_Y
  &=&  (Y_2^{UL})_{ij}\epsilon_{\alpha\beta}\ts
    \bar{u}_{R a  i}\tts R_{2}^{\tts a \alpha}\ell_{Lj}^{\ts\beta}
  + (Y_2^{EQ})_{ij}\ts
    \bar{e}_{Ri}\tts R_{2 a \alpha}^{\tts *}\ts q_{Lj}^{ a \alpha}
      + (Y_6^{DU})_{ij}\ts
    \bar{d}_{R a  i}\ts (\eta^A)^{ a  b}
    S_{6}^{A}\ts u_{R b j}^c \nonumber\\
  &+&\frac{(Y_6^{QQ})_{ij}}{2}\ts\epsilon_{\alpha\beta}\ts
    \bar{q}_{Li}^{\ts c\ts a \alpha} (\eta^A)_{ a  b}\ts
    S_{6}^{A*}\tts q_{Lj}^{ b\beta}
   \hc\,,
\label{eq:LYbelowGUT0}
\eea
where $\epsilon_{\alpha\beta}$ is an antisymmetric tensor with $\epsilon^{12} = \epsilon_{12} = 1$, $a, b = 1,2,3$ are SU(3) indices and $\alpha, \beta = 1, 2$ are SU(2) indices. Recall that the SU(5) fields can be decomposed into SM fields as follows:
\begin{equation}
\mathbf{10} \to (q_L, u^c_R, e^c_R), \quad \mathbf{\overline{5}} \to (d^c_R, \ell_L), \quad \mathbf{5}_H \to (H_u, T), \quad \mathbf{\overline{5}}_H \to (H_d, \overline{T}),
\end{equation}
where \(q_L\) and \(\ell_L\) are the left-handed quark and lepton doublets in the SM, \(u^c_R\), \(d^c_R\), and \(e^c_R\) are their right-handed charge conjugate counterparts, 
\( H_u \) and \( H_d \) represent the up- and down-type Higgs doublets, and \( T \) and \( \overline{T} \) are heavy color-triplet fields. The symmetric $3\times3$ matrices,  $\eta^{A}$, are as defined in~\cite{Han:2009ya} -- they play a key role in encoding the structure of interactions between the various fields. 
The couplings \( Y_6^{QQ} \) that appear in Eq.~(\ref{eq:LYbelowGUT0}) form an antisymmetric matrix over the generation indices \( i, j \). This antisymmetry of \( Y_6^{QQ}\) arises naturally from the GUT-scale matching conditions of the couplings, which impose the following relationship:  
\begin{equation}
Y_6^{QQ} = -\frac{1}{\sqrt{2}} Y_{45}^U,
\end{equation}  
where \( Y_{45}^U \) represents the Yukawa coupling associated with the \( \mathbf{45}_H \) representation in SU(5).  

The quark and lepton fields in Eq.~(\ref{eq:LYbelowGUT0}) were written in the flavor basis. After transforming to the mass basis, we can rewrite the flavor-basis quark and lepton fields as follows.
\begin{align}
q_{Li} = \begin{pmatrix} 
        (U_L)_{ij}\hat{u}_{Lj} \\
        (D_L)_{ij}\ts \hat{d}_{Lj}
        \end{pmatrix},~
u_{Ri} = (U_R)_{ij}\hat{u}_{Rj}\,,~
d_{Ri} = (D_R)_{ij}\hat{d}_{Rj}\,,~
\ell_{Li} = \begin{pmatrix}
           \hat{\nu}_{Li} \\
           \hat{e}_{Li}
           \end{pmatrix},~
e_{Ri} = \hat{e}_{Ri}\,,~ \label{eq:SMfermions}
\end{align}
where $\hat{x}$ denotes the mass eigenstate corresponding to the flavor-basis field $x$, $U_{L(R)}, D_{L(R)}$ are the left-handed (right-handed) up- and down-type quark mixing matrices, and $V_{\text{CKM}}=U_L^\dagger D_L$ represents the Cabibbo-Kobayashi-Maskawa (CKM) matrix \footnote{Note that in principle the neutrino fields also get rotated by the Pontecorvo-Maki-Nakagawa-Sakata (PMNS) matrix to describe neutrino masses and mixings. However, in this work, we will treat the neutrinos as massless. Furthermore, since the PMNS matrix is a unitary matrix and neutrino flavors are not detected, there is no observable effect of including the PMNS matrix. We, therefore, leave it out of the discussion.}. 

Expanding the Lagrangian in Eq.~(\ref{eq:LYbelowGUT0}) in terms of fermion fields in the mass basis using Eq.~(\ref{eq:SMfermions}), we obtain
\bea -\mathcal{L}_Y &=& (Y_2^{UL})_{ij}\left[\left(\ts\bar{\hat{u}}_{R a  i}\tts R_{2}^{\tts a  1} \hat{e}_{Lj}\right) - \left(\ts\bar{\hat{u}}_{Rai}\tts R_{2}^{\tts a  2} \hat{\nu}_{Lj}\right)\right] + (Y_6^{DU})_{ij}\ts\bar{\hat{d}}_{R a  i}\ts (\eta^A)^{ a  b}S_{6}^{A}\ts \hat{u}_{R b j}^c \nonumber\\
&& +~\frac{1}{2} (Y_6^{QQ})_{ij}\left[\ts \ts\bar{\hat{u}}_{Li}^{\ts c\ts a} (\eta^A)_{ a  b}\ts S_{6}^{A*}\tts (V_{\text{CKM}})_{jk}\ts\hat{d}_{Lk}^{ b} - \ts \ts (V_{\text{CKM}})^T_{ki}\ts\bar{\hat{d}}_{Lk}^{\ts c\ts a }(\eta^A)_{ a  b}\ts S_{6}^{A*}\tts \hat{u}_{Lj}^{\ts b}\right] \nonumber\\
&& +~(Y_2^{EQ})_{ij}\left[\ts\bar{\hat{e}}_{Ri}\tts R_{2 a  1}^{\tts*}\ts\hat{u}^{\tts a }_{Lj} + \ts\bar{\hat{e}}_{Ri}\tts R_{2 a  2}^{\tts *}\ts (V_{\text{CKM}})_{jk}\ts \hat{d}^{\tts a }_{Lk}\right], \hc\label{eq:LYbelowGUT} 
\eea
where, without the loss of generality, we have absorbed the mixing matrices in the Yukawa couplings through the following reparametrizations:
$Y_2^{UL} \to U_R^\dagger Y_2^{UL},~ Y_6^{DU} \to D_R^\dagger  Y_6^{DU} U_R$, 
$Y_6^{QQ} \to U_L^T  Y_6^{QQ} U_L,~ Y_2^{EQ} \to  Y_2^{EQ} U_L$. We will use the Particle Data Group (PDG) phase
convention~\cite{Chau:1984fp,ParticleDataGroup:2020ssz} when computing the CKM matrix elements.

The diquark and leptoquark couplings, $Y_6^{QQ,DU}$ and $Y_2^{UL,EQ}$, in the above Lagrangian can be constrained using experimental limits on FCNC processes, such as \(b\to s(d) \gamma\). This is because the diquark and leptoquark fields can replace the \(W\) boson in loop diagrams analogous to the SM penguin and box diagrams. After integrating out the heavier scalar fields, \(S_6\) and \(R_2\), one can compute their effects on effective operators. 
Focusing on the electromagnetic and chromomagnetic dipole operators,
\bea
O_7 &=& \frac{e}{16\pi^2} m_b(\mu) (\bar{q}_L\sigma_{\mu\nu}b_R) F^{\mu \nu}\,,  \quad  \quad O_8 ~=~ \frac{g_s}{16\pi^2} m_b(\mu) (\bar{q}_L T^a \sigma_{\mu\nu} b_R) G^{a \mu
\nu} \; , \\
\tilde{O}_7 &=& \frac{e}{16\pi^2} m_b(\mu) (\bar{q}_R\sigma_{\mu\nu}b_L) F^{\mu \nu}\,,  \quad  \quad \tilde{O}_8 ~=~ \frac{g_s}{16\pi^2} m_b(\mu) (\bar{q}_R T^a \sigma_{\mu\nu} b_L) G^{a \mu
\nu} \; , \label{operatorchiral}
\eea
one can readily show that \(S_6\) not only introduces new contributions to both \(O_7\) and \(O_8\) but also generates additional contributions to their chirality-flipped counterparts \(\tilde{O}_7\) and \(\tilde{O}_8\). The Wilson coefficients (WCs) corresponding to these operators obtained by integrating out \(S_6\) are given as,
\bea 
C_{7}^{ S_{6}}&=& \frac{
v^2}{4 \lambda_{tq} m_{ S_{6}}^2} \sum_{j,A,D=1}^{3}~ (Y^{QQ}_6)^*_{j
A} V^{CKM*}_{A q} (Y^{QQ}_6)_{j D} V^{CKM}_{D b} f_7(y_{j}) \nonumber\\
&+&\frac{
v^2}{2\lambda_{tq} m_{ S_{6}}^2} \frac{1}{m_b} \sum_{j,A=1}^{3}~
m_{u_{j}} (Y^{QQ}_6)^*_{j A} V^{CKM*}_{A q} (Y^{DU}_6)^*_{bj}
\tilde f_7(y_{j})  \, , \label{wilson3LOf1} \\ 
 C_{8}^{ S_{6}}&=&  \frac{ v^2}{4\lambda_{tq} m_{ S_{6}}^2}
\sum_{j,A,D=1}^{3}~
 (Y^{QQ}_6)^*_{j A} V^{CKM*}_{A
q} (Y^{QQ}_6)_{j D} V^{CKM}_{D b} f_8(y_{j}) \nonumber \\
&+&\frac{ v^2}{2\lambda_{tq}
m_{ S_{6}}^2} \frac{1}{m_b} \sum_{j,A=1}^{3}~ m_{u_{j}}
(Y^{QQ}_6)^*_{j A} V^{CKM*}_{A q} (Y^{DU}_6)^*_{bj}\tilde
f_8(y_{j}) \, ,\label{wilson3LOf2} \\
\tilde C_{7}^{ S_{6}}&=& \frac{
v^2}{\lambda_{tq} m_{ S_{6}}^2} \sum_{j=1}^{3}~ (Y^{DU}_6)^*_{bj}
(Y^{DU}_6)_{qj} f_7(y_{j})\nonumber\\
&+&\frac{ v^2}{2\lambda_{tq} m_{ S_{6}}^2}
\frac{1}{m_b}  \sum_{j,A=1}^{3}~ m_{u_{j}} (Y^{QQ}_6)_{j A}
V^{CKM}_{A b} (Y^{DU}_6)_{qj} \tilde f_7(y_{j}) \, , \label{wilson3LOf3} \\
\tilde C_{8}^{ S_{6}}&= &  \frac{ v^2}{\lambda_{tq} m_{ S_{6}}^2} \sum_{j=1}^{3}~
(Y^{DU}_6)^*_{bj}  (Y^{DU}_6)_{qj} f_8(y_{j}) \nonumber\\
&+&\frac{
v^2}{2\lambda_{tq} m_{ S_{6}}^2} \frac{1}{m_b} \sum_{j,A=1}^{3}~
m_{u_{j}} (Y^{QQ}_6)_{j A} V^{CKM}_{A b} (Y^{DU}_6)_{qj}\tilde
f_8(y_{j})\, , \label{wilson3LOf4}
\eea
where $v = (246/\sqrt{2})$ GeV is the SM Higgs VEV, $y_j=m_{u_j}^2/m_{S_{6}}^2$, and $f_{7,8}(y_{j})$ and $\tilde f_{7,8}(y_{j})$ are loop functions that take the following explicit forms \cite{Crivellin:2018qmi,Crivellin:2023saq}. 
\bea \label{c7xyetc} 
f_7(y_{j}) &=& \frac{1}{72} \, \left[
  \frac{4y_j^3-9 y_j^2+5-6 y_j (y_j-2)\ln y_j}{(y_j-1)^4} \right] \, ,\\
\tilde f_7(y_{j})  &=& \frac{1}{12} \, \left[ \frac{3y_j^2-8y_j+5-2(y_j-2) \ln y_j}{(y_j-1)^3} \right]\,, \\
f_8(y_{j}) &=& \frac{1}{24} \, \left[
\frac{-y_j^3-9y_j^2+9y_j+1+6 y_j (y_j+1) \ln y_j}{(y_j-1)^4}\right]\, , \\
\tilde f_8(y_{j}) &=& \frac{1}{2} \, \left[ \frac{-2y_j+2+(y_j+1)
\ln y_j}{(y_j-1)^3} \right] \, .
\eea

Focusing specifically on $b\to s\gamma$, assuming a TeV scale mass for the diquark, and plugging in numerical values of measured parameters such as the CKM matrix elements, we can expand sums in Eqs.~(\ref{wilson3LOf1})--(\ref{wilson3LOf4}) in terms of products of Yukawa couplings. We find that the dominant contributions to the WCs come from the two products $(Y^{DU}_6)_{33}(Y^{QQ}_6)_{32}$ and $(Y^{QQ}_6)_{33}(Y^{DU}_6)_{23}$, while the contributions from all other products of two Yukawas are subdominant. Following Refs.~\cite{Arnan:2019uhr, Misiak:2015xwa}, one can show that experimental bounds on $b\to s\gamma$ can be used to constrain $C_{7,8}$ and $\tilde C_{7,8}$ \footnote{The observable that has the highest sensitivity to new physics in $b\to s\gamma$ is the branching ratio ${\cal B}(B\to X_s\gamma)$. The SM prediction for this branching ratio was calculated at the next-to-next-to-leading order in Ref.~\cite{Misiak:2015xwa}. New physics effects can be captured by rescaling the SM prediction with functions that depend on new physics Wilson Coefficients \cite{Hurth:2003dk}. On the other hand, the CP- and isospin-averaged experimental value for this observable is reported by HFLAV~\cite{HFLAV:2022esi}.}. In turn, these can be used to put bounds on products of Yukawa couplings in our model. We find that the most stringent of these bounds is $|(Y^{DU}_6)_{33} (Y^{QQ}_6)_{32}| \lesssim 2.7\times 10^{-3}$, while less stringent bounds apply to all other products of two Yukawa couplings. On the other hand, the product $(Y^{DU}_6)_{33} (Y^{QQ}_6)_{32}$ does not contribute to the $B$ anomalies we discuss in this paper. Furthermore, we find that similar, albeit slightly weaker, constraints on products of Yukawa couplings can also be obtained from $b\to d\gamma$. Consequently, it is still possible to use the $S_6$ diquark in our model to analyze the $B$ anomalies without running afoul of constraints from $b\to s(d)\gamma$.

We now turn to the leptoquark, \(R_2\). The Wilson coefficients corresponding to \(O_{7,8}\) and \(\tilde{O}_{7,8}\) obtained by integrating out \(R_2\) can be summarized as follows.
\bea 
\label{wilson3LO1} 
C_{7}^{R_{2 a2}}&=& \frac{
v^2}{9 \lambda_{tq} m_{ R_{2 a2}}^2} \sum_{j,k,\ell=1}^{3}~ (Y^{EQ}_2)^*_{j
k} V^{CKM*}_{k q} (Y^{EQ}_2)_{j \ell} V^{CKM}_{\ell b}  \, ,\\ C_{8}^{ R_{2 a2}}&=& - \frac{ v^2}{12\lambda_{tq} m_{ R_{2 a2}}^2}
\sum_{j,k,\ell=1}^{3}~
 (Y^{EQ}_2)^*_{j k} V^{CKM*}_{k
q} (Y^{EQ}_2)_{j \ell} V^{CKM}_{\ell b}  \, ,\\
\tilde C_{7}^{ R_{2 a2}}&=&\tilde C_{8}^{ R_{2 a2}}=0\,,
\eea
where we have neglected the mass of the charged lepton running in the loop compared to the leptoquark mass. Here we find that, for $|(Y^{EQ}_2)^*_{ij} (Y^{EQ}_2)_{k \ell}| \lesssim 10^{-1}$, leptoquark contributions to these Wilson coefficients are small compared to their SM values. Thus, constraints on the leptoquark Yukawa couplings appear to be even weaker than those on the diquark Yukawa couplings.

Other flavor-violating processes, such as \(B - \bar{B}\), \(K - \bar{K}\), and \(D - \bar{D}\) mixing can receive new contributions from box diagrams with the diquark or leptoquark running in the loop. We follow Ref.~\cite{Ezzat:2022gpk}, which implements bounds corresponding
to ${\cal O}(10\text{--}20\%)$ deviations in $\Delta M_{B_{s}}$ and $\Delta M_{B_{d}}$ allowed by HFLAV results~\cite{HFLAV:2022esi} and theory estimates from Ref~\cite{DiLuzio:2019jyq}. However, in addition to being loop-suppressed, these new contributions are proportional to products of four new Yukawa couplings associated with the diquark or leptoquark. Since each Yukawa coupling can be modified independently, obtaining meaningful constraints on any one of them from meson mixing is not possible. Moreover, in the following sections, we focus on processes that receive new contributions from the diquark and/or leptoquark states involving only products of two new Yukawa couplings. Consequently, meson-mixing constraints on products of four new Yukawa couplings do not affect our results.

We have introduced a non-minimal SU(5) model with an \(R_2\) leptoquark and an \(S_6\) diquark. Both the diquark and the leptoquark are heavy. Additionally, numerous Yukawa couplings parameterize the interactions of the SM particles with \(R_2\) and \(S_6\). We have shown that, with couplings ranging from $10^{-1}$ to $10^{-2}$ or smaller, one can avoid experimental bounds from FCNC processes such as $b\to s(d)\gamma$ and meson mixing. In what follows, we will discuss the effects of the diquark and leptoquark on the anomalies in $B$ decays.

\section{SU(5) diquark contributions to $\btopiK$}

In this section, we consider the contributions of the diquark to $\btopiK$ decays and discuss a path for the resolution of the $\btopiK$ puzzle. To understand the puzzle, it is advantageous to express the $\btopiK$ decay amplitudes in terms of topological flavor-flow amplitudes \cite{Gronau:1994rj, Gronau:1995hn, Fleischer:1997um, Gronau:1997an, Neubert:1998pt, Neubert:1998jq, Gronau:2006xu} commonly referred to as the penguin ($P$), color-favored tree ($T$), color-suppressed tree ($C$), and annihilation ($A$) amplitudes. The penguin amplitudes are further categorized as the QCD penguin amplitude $P$, electroweak penguin (EWP) amplitude $P_{\rm EW}$, and color-suppressed EWP amplitude $P^C_{\rm EW}$ originating respectively from the QCD penguin and electroweak penguin operators in the dimension-6 hadronic effective Hamiltonian \cite{Buchalla:1995vs,Gronau:1998fn}. In general, individual flavor-flow amplitudes carry a strong phase and a weak phase and the relative phases between different amplitudes become important at the level of decay observables.

The decay amplitudes for $\bar{B}^0\to K^-\pi^+, B^-\to\bar{K}^0\pi^-, \bar{B}^0\to\bar{K}^0\pi^0$, and $B^-\to K^{-}\pi^0 $ are expressed as \cite{Chau:1990ay, Gronau:1994rj, Gronau:1995hn, Gronau:2005kz},
\bea
\mathcal{A}(\bar{B}^{0}\to K^{-}\pi^{+}) &=& -\lambda_{u} \left(P_{uc}+T\right) - \lambda_{t} \left(P_{tc}+\frac23 P^{C}_{EW}\right)\,, \\
\mathcal{A}(B^{-} \to \bar{K}^{0}\pi^{-}) &=& \lambda_{u} \left(P_{uc}+A \right)+\lambda_{t} \left(P_{tc}-\frac{1}{3}P^{C}_{EW}\right)\,, \\
\sqrt{2}\mathcal{A}(\bar{B}^{0}\to \bar{K}^{0}\pi^{0}) &=& \lambda_{u} \left(P_{uc}-C\right) + \lambda_{t} \left(P_{tc}-P_{EW}-\frac{1}{3}P^{C}_{EW}\right) \,, \\
\sqrt{2}\mathcal{A}(B^{-}\to K^{-}\pi^{0}) &=& -\lambda_{u} \left(P_{uc}+T+C+A\right) - \lambda_{t} \left(P_{tc}+P_{EW}+\frac{2}{3}P^{C}_{EW}\right)\,, \label{Bto1P}
\eea
where $\lambda_{q} = V^{*}_{qs}V_{qb}^{}$, $P_{uc} = P_u - P_c$, and $P_{tc} = P_t - P_c$. The subscript ($u,c,t$) on the penguin amplitude denotes the up-type quark running in the loop. 

Now, a combination of effects such as the hierarchy of CKM elements, loop suppression in penguins, and other dynamical effects leads to the following rough hierarchy in the sizes of the flavor-flow amplitudes \cite{Gronau:1995hm},
\begin{align}
\left|\lambda_{t}P_{tc}\right| \, > \, \left|\lambda_{u}T\right|\, > \, 
\left|\lambda_{u}C\right| \, > \, \left|\lambda_{u}A\right|\,,\,  \left|\lambda_{u}P_{uc}\right|\,.
\label{eq:hier}
\end{align}
One can estimate ratios such as $|C/T|$ and $|P_{tc}/T|$ using pQCD \cite{Keum:2000ms} or QCD factorization \cite{Beneke:2000ry,Beneke:2001ev}, however, there is little consensus on the results. For example, the ratio $|P_{tc}/T|$ obtained using QCD factorization differs from its pQCD estimate of $\sim 0.1$ \cite{Keum:2000ms}. The ratio $|C/T|$ is found to be $\sim 0.2$ \cite{Beneke:2001ev,Bell:2007tv,Bell:2009nk,Beneke:2009ek,Bell:2015koa,Bell:2020qus}, or $\sim 0.5$ \cite{Li:2009wba}, or even approaching unity~\cite{Bauer:2005kd,Huitu:2009st}, depending on the scenario. The ratio $|A/T|$ is expected to be small as it is suppressed by $f_{B}/m_{B}\sim 0.05$, the ratio of the $B$-meson decay constant to its mass. In the SM, the EWP amplitudes can be estimated using the following relationship between the tree and EWP amplitudes \cite{Gronau:1998fn}.
\beq
P_{EW} \pm P_{EW}^{C}=-\frac{3}{2}\, \frac{C_{9} \pm C_{10}} {C_{1} \pm C_{2}}\, (T \pm C)\,.
\eeq
The above relationship uses $SU(3)$-flavor symmetry of the dimension-6 weak Hamiltonian \cite{Buchalla:1995vs,Gronau:1998fn}. Using numerical values of the Wilson coefficients $C_{1,2}$ and $C_{9,10}$ to the leading-log order at the $m_{b}$ scale~\cite{Buchalla:1995vs}, one gets
\beq
\label{T-PEW reln}
P_{EW}\sim \kappa T\,,\ \ \qquad
P_{EW}^{C}\sim \kappa C\,,
\eeq
to a good approximation, where 
\beq
\kappa=-\frac{3}{2}\, \frac{C_{9}+C_{10}}{C_{1}+C_{2}}\simeq -\frac{3}{2}\, \frac{C_{9}-C_{10}}{C_{1}-C_{2}}\simeq 
0.0135\pm 0.0012\,.
\label{eq:kappa-def}
\eeq 

Ignoring the small amplitudes $C, A,$ and $P_{uc}$ one finds that the direct-CP asymmetries in $\bar{B}^0\to K^-\pi^+$ and $B^-\to K^-\pi^0 $ are identical, i.e.~\cite{Gronau:1998ep}
\beq
\label{simpleacp}
	A_{\rm CP}(\bar{B}^{0}\to K^- \pi^{+})=A_{\rm CP}(B^{-}\to K^-\pi^{0}) ~.
\eeq
This can be seen as follows. Once the small amplitudes are neglected, the CP asymmetries in both $\bar{B}^0\to K^-\pi^+$ and $B^-\to K^-\pi^0$ can only come from the interference between $T$ and $P_{tc}$, as they have different weak and strong phases. Despite the difference in relative weak phases between the $P_{EW}$ and $T$ terms, the interference between these two terms does not generate a direct-CP asymmetry in $B^-\to K^-\pi^0$ as the strong phase difference between them vanishes in the SM, as can be seen from  Eq.~\eqref{eq:kappa-def}. 

In experiments, a deviation from the relation given in Eq.~(\ref{simpleacp}) is measured by the quantity $\Delta A_{\text{CP}}$,
\begin{align}\label{eq:delACP}
\Delta A_{\rm CP}=A_{\rm CP}(B^{-}\to K^{-}\pi^{0})-A_{\rm CP}(\bar{B}^{0}\to K^{-}\pi^{+}),
\end{align}
The global experimental average of $\Delta A_{\text{CP}}$,
\begin{align}
\Delta A_{\rm CP}\vert_{\text{avg}}=0.110\pm 0.012, 
\end{align}
is based on measurements from  Belle~\cite{Duh:2012ie}, Belle-II~\cite{Belle-II:2023ksq}, and LHCb~\cite{Aaij:2020wnj}.
In addition to the non-zero value of $\Delta A_{\rm CP}$, the relative sign of $A_{\text{CP}}$ in the two decay modes is in contradiction with the SM expectation. Including the small amplitudes that were ignored while arriving at Eq.~(\ref{eq:delACP}), one can construct a more theoretically robust sum-rule relationship \cite{Gronau:2005kz} connecting the branching ratios (${\cal B}$) and CP asymmetries in all four $\btopiK$ decay channels, expressed as \cite{Gronau:2005kz}, 
\bea \label{isospin-sumrule}
\Delta_{4} &=& A_{\rm CP}(\bar{B}^{0}\to K^-\pi^+) + A_{\rm CP}(B^-\to\bar{K}^0 \pi^-) \frac{{\cal B}(B^-\to K^0\pi^-) \tau_{B^0}}{{\cal B}(\bar{B}^0\to K^-\pi^+) \tau_{B^{-}}} \nonumber \\
&& -~A_{\rm CP}(B^-\to K^-\pi^0)\frac{2{\cal B}(B^-\to K^-\pi^0)\tau_{B^{0}}}{{\cal B}(\bar{B}^0\to K^-\pi^+)\tau_{B^{-}}} \nonumber \\
&& -~A_{\rm CP}(\bar{B}^0\to\bar{K}^0\pi^0)\frac{2{\cal B}(\bar{B}^0\to\bar{K}^0\pi^0)}{{\cal B}(\bar{B}^0\to K^-\pi^+)}\,,~~
\end{eqnarray}
where $\tau_{B^{-}}$ and $\tau_{\bar{B}^{0}}$ are
the lifetimes of the $B^{-}$ and $\bar{B}^{0}$ mesons, respectively. Eq.~(\ref{isospin-sumrule}) does not assume relative sizes of amplitudes or strong phases, however, one can show that $\Delta_4$ vanishes (up to a few percent) in the limit of a small relative strong phase between $T$ and $C$ and $A$ much smaller than $T$. A previous measurement of $\Delta_{4}=-0.270 \pm 0.132 \pm 0.060$ at Belle~\cite{Duh:2012ie} deviated from zero, albeit with large errors. A recent measurement of $\Delta_{4}=-0.03 \pm 0.13 \pm 0.04$ at Belle~II~\cite{Belle-II:2023ksq}, however, is in good agreement with zero. 

In light of these experimental observations, we adopt an approach in which new physics (NP) contributions to the $B\to K\pi$ decays improve the explanation of the available data. The contribution from NP is modeled in terms of the NP matrix elements $\langle K\pi \vert Q_{k}^{\text{NP}}\vert B\rangle$. Each matrix element has its own NP weak and strong phases. The complete amplitude for all four $B\to K\pi$ decays can be written using the flavor-flow amplitudes in the SM and these NP matrix elements. 

Starting from the Lagrangian describing the color sextet diquark-quark interaction in Eq.~\ref{eq:LYbelowGUT}, we find that integrating out the diquark $S_{6}^{A}$ leads to the effective Lagrangian contributing to $\Delta B= \vert \Delta S \vert = 1$ transitions at tree level. The relevant dimension-6 effective operators $\tilde Q^{ij}_{k}$ and their corresponding NP Wilson coefficients $\tilde C^{ij}_{k}$ mediating $b\to u s \bar{u}$ transitions relevant for our study can be expressed as,
\beq 
{\mathcal H}_{\eff}^{\text{NP}}= -\frac{G_{F}}{\sqrt{2}}\sum^6_{k=1}\big( Q^{ij}_k
C^{ij}_k+\tilde Q^{ij}_k \tilde C^{ij}_k\big)\,,
\label{effHam}
\eeq%
where
\bea 
Q^{ij}_1 &=& \big(\overline{ u^\alpha_i} \gamma^\mu P_L
b^\alpha \big) \big(\overline{ s^\beta} \gamma_\mu P_L u^\beta_j
\big) \,,\nonumber\\
Q^{ij}_2 &=& \big(\overline{ u^\alpha_i} \gamma^\mu P_L b^\beta
\big) \big(\overline{ s^\beta} \gamma_\mu P_L
u^\alpha_j\big) \,,\nonumber\\
Q^{ij}_3 &=&\big( \overline{ u^\alpha_i} P_L
b^\alpha\big)\big(\overline{ s^\beta} P_L u^\beta_j\big) \,, \nonumber\\
Q^{ij}_4 &=& \big(\overline{ u^\alpha_i} P_L b^\beta\big)\big(s^\beta P_L
\overline{ u^\alpha_j}\big) \,,\nonumber\\
Q^{ij}_5 &=&\big(\overline{ u^\alpha_i}  \sigma^{\mu\nu} P_L b^\alpha\big)\big(\overline{ s^\beta} \sigma_{\mu\nu} P_L u^\beta_j\big) \,, \nonumber\\
Q^{ij}_6 &=& \big( \overline{ u^\alpha_i} \sigma^{\mu\nu} P_L b^\beta\big)\big(s^\beta  \sigma_{\mu\nu} P_L \overline{u^\alpha_j}\big)\,.
\label{Q11}
\eea  
The operators $\tilde Q^{ij}_k$ can be obtained from $Q^{ij}_k$ ($i,j$ are the generation indices of the up-type quark), for $k=1,2,..,6$, by switching the chirality operator from $P_L = (1 - \gamma^5)/2$ to $P_R = (1 + \gamma^5)/2$.  

The  Wilson coefficients $C^{ij}_k$ corresponding to the operators $Q^{ij}_k$ can be expressed as,
\bea 
C^{ij}_1 &=&  C^{ij}_2 = - \frac{\sqrt{2}}{8 G_{F} \, m^2_{S_{6}}} \sum^3_{k,\ell=1}  (Y^{QQ*}_6)_{i \ell} V^{CKM*}_{\ell 2} (Y^{QQ}_6)_{jk} V^{CKM}_{k3}\,,\\
C^{ij}_3 &=&  C^{ij}_4 = \frac{\sqrt{2}}{4 G_{F} \,m^2_{S_{6}}}
\sum^3_{k=1} (Y^{DU}_6)_{2 i}
(Y^{QQ}_6)_{jk} V^{CKM}_{k3}\,, \\
C^{ij}_5 &=&  C^{ij}_6 = - \frac{\sqrt{2}}{16 G_{F} \, m^2_{S_{6}}}\sum^3_{k=1}  (Y^{DU}_6)_{2 i} (Y^{QQ}_6)_{jk} V^{CKM}_{k3}\,,
\label{BKpi}
\eea
where $m_{S_{6}}$ is the mass of the sextet diquark. On the other hand, the Wilson coefficients $\tilde{C}^{ij}_k$, corresponding to the operators $\tilde{Q}^{ij}_k$ with
$k=1,2,\ldots,6$, are given as,
\bea
\tilde C^{ij}_1 &=&  \tilde C^{ij}_2=- \frac{\sqrt{2}}{2 G_{F} \, m^2_{S_{6}} } (Y^{DU}_6)_{2 i} (Y^{DU}_6)^*_{3j} \,, \nonumber\\
\tilde C^{ij}_3 &=&  \tilde C^{ij}_4 =  \frac{\sqrt{2}}{4 G_{F} \,m^2_{S_{6}}} \sum^3_{k=1} (Y^{QQ}_6)^*_{i k} V^{CKM*}_{k 2} (Y^{DU}_6)^*_{3j}  \,\nonumber\\
\tilde C^{ij}_5 &=&\tilde C^{ij}_6 = - \frac{\sqrt{2}}{16 G_{F}\,m^2_{S_{6}}} \sum^3_{k=1} (Y^{QQ}_6)^*_{i k} V^{CKM*}_{k 2} (Y^{DU}_6)^*_{3j} \,. 
\eea

Using the effective Hamiltonian listed in Eq.~(\ref{effHam}), processes in the factorization approximation, there are two independent NP contributions to the $\btopiK$ processes. With the help of hadronic matrix elements provided in Appendix \ref{app:inputs}, the $B\to K \pi$ NP decay amplitudes can be expressed as,
\bea 
{\mathcal A}^{\phi}\big(\bar{B}^0\to \pi^+ K^-\big) &=& \langle \pi^{+}K^{-}\vert \mathcal{H}^{\text{NP}}_{\eff} \vert \bar{B}^{0}\rangle \nonumber\\
&=& -\frac{i\,G_F}{4\sqrt{2}}\left(1+\frac{1}{N_c}\right) f_K\left(m_B^2-m_\pi^2\right) F^{{\bar B}\to \pi}_0(q^2=m_{K}^{2}) \nonumber \\
&& \left[\big(C^{uu}_1- \tilde{C}^{uu}_1 \big) - \frac{ m_K^2}{(m_u+m_{s})(m_{b}-m_{u})} \big( C^{uu}_3 - \tilde{C}^{uu}_3\big)\right] \,, \\
{\mathcal A}^{\phi} \big(\bar{B}^0\to \pi^0 \bar{K}^0\big) &=& \langle \pi^{0}\bar{K}^{0} \vert \mathcal{H}^{\text{NP}}_{\eff} \vert \bar{B}^{0}\rangle \nonumber \\
&=& \frac{i\,G_F}{16}\left(1+\frac{1}{N_c}\right) f_\pi \left(m_B^2-m_K^2\right) F^{{\bar B}\to K}_0(q'^2=m_{\pi}^{2}) \nonumber\\
&& \left[\big(C^{uu}_1- \tilde{ C}^{uu}_1 \big) + \frac{ m_\pi^2}{4 m_u (m_{b}-m_{s})} \big( C^{uu}_3- \tilde{C}^{uu}_3\big)\right] \,.
\eea
The $B^-\to\pi^-\bar{K}^{0}$ decay amplitude does not receive any NP contribution in the factorization approximation. The fourth amplitude, ${\mathcal A}^{\phi} \big(B^-\to \pi^0 K^-\big)$, is related to the first three through the isospin relation,
\begin{align}
\sqrt{2}\mathcal{A} (\bar{B}^{0}\to
\pi^{0}\bar{K}^{0})-\sqrt{2}\mathcal{A} (B^{-}\to \pi^{0}
K^{-})=\mathcal{A}(B^{-}\to
\pi^{-}\bar{K}^{0})-\mathcal{A}(\bar{B}^{0}\to \pi^{+}K^{-})\,.
\end{align}
In particular, since ${\mathcal A}^{\phi} \big(B^-\to \pi^- \bar{K}^{0}) = 0$,
\begin{equation}
{\mathcal A}^{\phi} \big(B^-\to \pi^0 K^-\big) ~=~ {\mathcal A}^{\phi} \big({\bar B}^0\to \pi^0 \bar{K}^0) + \frac{1}{\sqrt{2}}\,{\mathcal A}^{\phi} \big(\bar{B}^0\to \pi^+ K^-\big)\,.
\end{equation}
This ensures that only three combinations of NP amplitudes can be extracted from the data. To make the correspondence between NP amplitudes and hadronic matrix elements more transparent, we parametrize the $\btopiK$ decay amplitudes as follows. 
\bea
\mathcal{A}(\bar{B}^{0}\to \pi^{+}K^{-}) &=&-\lambda_{u}T-\lambda_{t}(P_{tc}+\frac{2}{3}P_{C}^{EW})-T^{\phi}\,, \label{BtoKPNP1} \\
\mathcal{A}(B^{-}\to \pi^{-}\bar{K}^{0})&=&\lambda_{t}(P_{tc}-\frac{1}{3}P_{C}^{EW})\,, \label{BtoKPNP2} \\
\sqrt{2}\mathcal{A} (\bar{B}^{0}\to \pi^{0}\bar{K}^{0})&=&-\lambda_{u}C+\lambda_{t}(P_{tc}-\frac{1}{3}P_{C}^{EW}-P_{EW})-C^{\phi}\,, \label{BtoKPNP3} \\
\sqrt{2}\mathcal{A} (B^{-}\to \pi^{0}
K^{-})&=&-\lambda_{u}(T+C)-\lambda_{t}(P_{tc}+\frac{2}{3}P_{C}^{EW}+P_{EW})
- (T^{\phi}+C^{\phi})\,, \label{BtoKPNP4}
\eea
where the NP amplitudes $T^{\phi}$ and $C^{\phi}$ are defined as,
\bea
\label{WC-diag correspondence}
T^{\phi}  &\equiv& \frac{iG_F}{4\sqrt{2}}
\left(1+\frac{1}{N_c}\right) f_K \big(m_B^2-m_\pi^2\big) F^{\bar B\to \pi}_0(q^2=m_{K}^{2}) \nonumber \\
&&\hspace{2truecm} \left[
 \big(C^{uu}_1- \tilde{ C^{uu}_1} \big)
- \frac{ m_K^2}{(m_u+m_{s})(m_{b}-m_{u})} \big( C^{uu}_3- \tilde
C^{uu}_3\big)\right] \,, \\
C^{\phi} &\equiv& -\frac{iG_F}{16}\left(1+\frac{1}{N_c}\right) f_\pi
\big(m_B^2-m_K^2\big)F^{\bar B\to K}_0(q'^2=m_{\pi}^{2}) \nonumber\\
&&\hspace{2truecm} \left[
 \big(C^{uu}_1- \tilde C^{uu}_1 \big)
+  \frac{ m_\pi^2}{4 m_u (m_{b}-m_{s})} \big( C^{uu}_3- \tilde
C^{uu}_3\big)\right]\,. 
\eea

We expect the strong phases associated with $T^{\phi}$ and $C^{\phi}$ to be very small \cite{Datta:2004re} as self-rescattering of these NP amplitudes is intrinsically feeble \cite{Datta:2004re}. However, there may be NP weak phases that contribute to these amplitudes through complex Wilson coefficients. More explicitly, the Wilson coefficient combinations $\big(C^{uu}_1- \tilde{ C^{uu}_1} \big)$ and $\big(C^{uu}_3- \tilde{ C^{uu}_3} \big)$ can individually carry weak phases $\phi_{1}$ and $\phi_{3}$ respectively. Interestingly, even if one of these weak phases vanishes, the overall weak phases for the new physics amplitudes $T^{\phi}$ and $C^{\phi}$ may differ from each other. Therefore, we investigate the following three scenarios: Scenario IIa: $\phi_{1}\neq 0$, $\phi_{3}\neq 0$, Scenario IIb: $\phi_{1}= 0$, $\phi_{3}\neq 0$, and Scenario IIc: $\phi_{1}\neq 0$, $\phi_{3}=0$.

\begin{table}[h!]
    \centering
    \footnotesize
        \begin{tabular}{|c|c|c|c|}
  \hline \hline
            Modes& Avg. ${\cal B}~~[10^{-6}]$  & Avg. $A_{\rm CP}$  &$S_{\rm CP}$\\
            \hline
            $\bar{B}^{0}\to \pi^{+}K^{-}$     &       $20.0\pm 0.4$           &         $-0.0831\pm 0.0031$ &   \\
            \hline
            $B^{-}\to \pi^{0}K^{-}$     &       $13.2\pm 0.4$             &         $0.027\pm 0.012 $   &   \\
            \hline
            $B^{-}\to \pi^{-}\bar{K}^{0}$     &       $23.9 \pm 0.6$            &         $-0.003 \pm 0.015 $ &   \\
            \hline
            $\bar{B}^{0}\to \pi^{0}\bar{K}^{0}$     &       $10.1\pm 0.4$             &         $0.00\pm 0.08 $ &   $0.64 \pm 0.13$ \\
   \hline\hline
        \end{tabular}
    \caption{\small Experimental inputs used in this work taken from PDG \cite{ParticleDataGroup:2024cfk}.}
    \label{Tab:avg data}
\end{table}
We use the PDG averaged data \cite{ParticleDataGroup:2024cfk} summarized in Table~\ref{Tab:avg data}. The data consists of the branching ratios and direct CP asymmetries for the four $B\to K\pi$ modes and the mixing-induced CP asymmetry, $S_{\rm CP}$, measured for the $\bar{B}^{0}\to\pi^0 \bar{K}^{0}$ decay from BaBar, Belle, Belle~II and LHCb~\cite{Belle-II:2023ksq, Aaij:2020wnj}. We perform five different fits to obtain the magnitudes and strong phases of the amplitudes in Eqs.~(\ref{BtoKPNP1}) -- (\ref{BtoKPNP4}).

First, we perform the SM fit, setting the NP amplitudes to zero. This fit uses five free fit parameters: the three magnitudes $P_{tc}, |T|, |C|$, and the two relative strong phases $\delta_{T},\, \delta_{C}$. The parameter $\kappa$ is treated as a prior around its central value. We have defined the strong phases of $T$ and $C$ relative to $P_{tc}$. The absolute phase of $P_{tc}$ is set to zero in this convention \cite{Kim:2007kx}. The magnitudes of the CKM elements $\vert V_{ub} \vert$, $\vert V_{us} \vert$, $\vert V_{tb} \vert$, $\vert V_{ts} \vert$, and the CKM angles $\beta$ and $\gamma$, also enter the analysis as SM priors. For the values of these CKM parameters, we take their HFLAV averages \cite{HFLAV:2022esi} listed in Table~\ref{Tab:ckmavg}.
\begin{table}[h!]
\centering
\begin{tabular}{|c|c|}	
\hline
			Parameter& Value\\ 
			\hline
			$\vert V_{us}\vert$		&		$0.2245\pm 0.0008$		\\ \hline
            $\vert V_{ub}\vert$     &       $(3.82 \pm 0.24)\times10^{-3}$\\ \hline
            $\vert V_{ts}\vert$     &       $(41.5 \pm 0.9)\times10^{-3}$\\ \hline
            $\gamma$     &       $(65.9 \pm 3.3 \pm 3.5)^{\circ}$\\ \hline
             $\beta$     &       $(22.14 \pm 0.69 \pm 0.67)^{\circ}$\\
 \hline
\end{tabular}
\caption{Relevant theoretical CKM input parameters from HFLAV averages~\cite{HFLAV:2022esi}.}
\label{Tab:ckmavg}
\end{table}

The results of the SM fit (with NP amplitudes set to zero) are given in Scenarios Ia and Ib of Table~\ref{tab:tabSMfit}. Scenario Ia refers to a fit in which $|C/T| < 0.2$ -- here one finds a very poor fit. Relaxing this requirement results in a better fit presented in Scenario Ib. This fit has a somewhat acceptable $p$-value, however, the preferred $|C/T|$ value is larger $\sim 0.5$. Additionally, if the CKM angle $\gamma$ is set to the recently measured value of ${78.6^\circ}^{+7.2^\circ}_{-7.3^\circ}$ at Belle~II~\cite{Belle:2024knt}, we find that $|C/T|$ shifts toward an even larger value of $0.68$ with an increased $\Delta \chi^{2}/{d.o.f}$ and a reduction in the $p$-value.
\begin{table}[!htbp]
\begin{tabular}{|c|c|c|c|c|}
\hline
Scenario &$\chi^{2}/d.o.f$&$p$-value & Fit parameter & Fit value\\
\hline
 &  & &$P_{tc}$ & $-0.150\pm 0.001$ \\
\textbf{Ia}&4.2 & 0.002 & $\kappa$ & $0.0134\pm  0.001$ \\
SM fit: $|C/T|\leq 0.2$& &  &$\vert T \vert $ & $1.21\pm 0.15$\\
&  &  &$\vert C \vert $ & $0.24\pm 0.04$\\
 & & &$\delta_{T} $ & $3.36\pm 0.03$\\
 & &  &$\delta_{C} $ & $1.58\pm 0.26$\\
\hline \hline
 &  & &$P_{tc}$ & $-0.149\pm 0.001$\\
\textbf{Ib}&1.4 & 0.23 & $\kappa$ & $0.0134\pm 0.001$ \\
SM fit& &  &$\vert T \vert $ & $0.84\pm 0.20$\\
&  &  &$\vert C \vert $ & $0.41\pm 0.09$\\
 & & &$\delta_{T} $ & $3.48\pm 0.09$\\
 & &  &$\delta_{C} $ & $1.14\pm 0.45$\\
\hline \hline
\end{tabular}%}
\caption{Best fit values for magnitudes and relative strong phases of diagrammatic amplitudes in the SM. Here, $G_{F}/\sqrt{2}$ has been factored out of the magnitude of each diagram. Magnitudes are quoted in units of $(\text{GeV})^{3}$ and phases in radians. 
\label{tab:tabSMfit}}
\end{table}
\begin{table}[!htbp]
\begin{tabular}{|c|c|c|c|c|}
\hline
Scenario &$\chi^{2}/d.o.f$&$p$-value & Fit parameter & Fit value\\
\hline
& & \, & $P_{tc}$ & $-0.149\pm 0.001$ \\
& & \, & $\kappa$ & $0.0135\pm 0.001$ \\
\textbf{IIa}& 0.95  &0.41  &$\vert T \vert $ & $1.02\pm 0.45$\\ 
NP amplitudes & & &$\vert C \vert $ & $0.30\pm 0.08$\\
 $T^{\phi},C^{\phi}$& & &$\delta_{T} $ & $3.34\pm 0.08$\\
2 NP weak phase $\phi_{1}$ \& $\phi_{3}$ & &  &$\delta_{C} $ & $1.14\pm 1.15$\\
 &  & &$\vert C_{1}^{uu}-\tilde{C_{1}^{uu}}\vert$ &$0.00038\pm0.00074$ \\
 & & &$\vert C_{3}^{uu}-\tilde{C_{3}^{uu}} \vert$ & $0.018\pm 0.012$\\
 & & & $\phi_{1}$ & $4.17\pm 0.60$\\
  & & & $\phi_{3}$ & $1.10\pm 0.29$\\
\hline
& & \, & $P_{tc}$ & $-0.148\pm 0.001$ \\
& & \, & $\kappa$ & $0.0135\pm 0.001$ \\
\textbf{IIb}& 0.79 &0.53  &$\vert T \vert $ & $1.10\pm 0.46$\\ 
NP amplitudes & & &$\vert C \vert $ & $0.33\pm 0.15$\\
 $T^{\phi},C^{\phi}$& & &$\delta_{T} $ & $3.34\pm 0.08$\\
1 NP weak phase $\phi_{1}=0$ \& $\phi_{3}$ & &  &$\delta_{C} $ & $1.19\pm 1.18$\\
 &  & &$\vert C_{1}^{uu}-\tilde{C_{1}^{uu}}\vert$ &$0.00045\pm0.00016$ \\
 & & &$\vert C_{3}^{uu}-\tilde{C_{3}^{uu}} \vert$ & $0.013\pm 0.006$\\
  & & & $\phi_{3}$ & $1.21\pm 0.18$\\
  \hline
& & \, & $P_{tc}$ & $-0.149\pm 0.001$ \\
& & \, & $\kappa$ & $0.0135\pm 0.001$ \\
\textbf{IIc}& 1.84 &0.13  &$\vert T \vert $ & $1.24\pm 0.79$\\ 
NP amplitudes & & &$\vert C \vert $ & $0.48\pm 0.32$\\
 $T^{\phi},C^{\phi}$& & &$\delta_{T} $ & $3.37\pm 0.16$\\
1 NP weak phase $\phi_{3}=0$ \& $\phi_{1}$ & &  &$\delta_{C} $ & $0.91\pm 0.78$\\
 &  & &$\vert\tilde{C_{1}^{uu}}-\tilde{C_{1}^{uu}}\vert$ &$0.0026\pm 0.0040$ \\
 & & &$\vert C_{3}^{uu}-\tilde{C_{3}^{uu}} \vert$ & $0.0017\pm 0.0032$\\
  & & & $\phi_{1}$ & $0.73\pm 0.91$\\
\hline \hline 
\end{tabular}%}
\caption{Best fit values for magnitudes and relative strong phases of diagrammatic amplitudes including NP contributions. Magnitudes are quoted in units of $(\text{GeV})^{3}$ and phases in radians. 
\label{tab:tabNPfit}}
\end{table}

Using the values obtained for $P_{tc}, |T|$, and $\kappa$ in Scenario Ia as priors, in Scenarios IIa, IIb, and IIc, we add the two complex NP amplitudes, $T^{\phi}$ and $C^{\phi}$. We then obtain the best fit values of $\phi_{1},\, \phi_{3}$, $\vert C^{uu}_1- \tilde{ C^{uu}_1} \vert$, and $\vert C^{uu}_3- \tilde{ C^{uu}_3} \vert$ in Scenario IIa. In Scenario IIb, $\phi_1$ is set to zero and the best fit values of $\phi_{3}$, $\vert C^{uu}_1- \tilde{ C^{uu}_1} \vert$ and $\vert C^{uu}_3- \tilde{ C^{uu}_3} \vert$ are obtained. Finally, in Scenario IIc, setting $\phi_3 = 0$, we get the best fit values of $\phi_{1}$, $\vert C^{uu}_1- \tilde{ C^{uu}_1} \vert$ and $\vert C^{uu}_3- \tilde{ C^{uu}_3} \vert$. The results of all three NP scenarios are presented in Table~\ref{tab:tabNPfit}, where we have used the values for the decay constants \cite{ParticleDataGroup:2022pth}, $B\to\pi,K$ form factors \cite{Ball:2004ye}, and masses \cite{ParticleDataGroup:2022pth} as inputs.

These results constrain the Yukawa matrices $Y_{6}^{QQ}$ and $Y_{6}^{DU}$. For illustration, choosing $(Y_{6}^{QQ})_{13}$ and $(Y_{6}^{DU})_{21}$ to be the only non-zero entries in $Y_{6}^{QQ}$ and $Y_{6}^{DU}$ matrices, we find
\bea
C_{1}^{uu} &=& -\frac{\sqrt{2}}{8 G_{F}\, m^2_{S_{6}}}(Y^{QQ*}_6)_{13} V^{CKM*}_{32} (Y^{QQ}_6)_{13} V^{CKM}_{33}\,,~~~ \tilde{C}_{1}^{uu}~=~ 0 \,, \\
C_{3}^{uu} &=& \frac{\sqrt{2}}{4 G_{F} \,m^2_{S_{6}}}(Y^{DU}_6)_{2 1}
(Y^{QQ}_6)_{13} V^{CKM}_{33}\,,~~~  \tilde{C}_{3}^{uu} ~=~ 0\,. 
\eea
For a benchmark value of the sextet diquark mass, $m_{S_{6}} = 2$ TeV, the following relations between the elements of the $Y_{6}^{QQ}$ and $Y_{6}^{DU}$ matrices have to be satisfied:
\bea
\left|-0.00375\sum\limits^3_{k,\ell=1}  (Y^{QQ*}_6)_{1 \ell} V^{CKM*}_{\ell 2} (Y^{QQ}_6)_{1k} V^{CKM}_{k3} + 0.015 (Y^{DU}_6)_{2 1} (Y^{DU}_6)^*_{31} \right| &\simeq& 0.00045 \,,~~~~~~~ \\
\left|0.075\sum\limits^3_{k=1}(Y^{DU}_6)_{2 1}(Y^{QQ}_6)_{1k} V^{CKM}_{k3}-0.075(Y^{QQ}_6)^*_{1 k} V^{CKM*}_{k 2} (Y^{DU}_6)^*_{31} \right|  &\simeq& 0.013\,.~~~~~
\eea

Including NP contributions, we find,
\begin{align}
    A_{\text{CP}}(B^{0}\to \bar{K}^{0}\pi^{0})&=-0.077\pm 0.035,\nn\\
    \Delta_{4}&=-0.04\pm 0.14.
\end{align}

\section{SU(5) leptoquark contributions to the $B$ anomalies at tree level}

In this section, we explore the contributions of the $R_2$ leptoquark arising in non-minimal SU(5) to semileptonic $B$ decays. The relevant quark-level processes are $b \to c \ell \bar{\nu}_\ell$ and $b \to s \ell^+ \ell^-$. These decays proceed through a tree-level exchange of leptoquarks. In presence of NP, the effective Hamiltonian for $|\Delta c| = 1$ semileptonic $B$ decays can be written as, 
\bea
{\cal H}_{\eff} &=& \frac{4G_F}{\sqrt{2}}V_{cb}
\sum\limits_{i,j}^{1,2,3}\left[\sum\limits_{X,Y}^{L,R}\left\{(\delta^{XL}\delta^{YL} + g_V^{XYij})({\bar c}\gamma_\mu P_X b)({\bar\ell_j}\gamma_\mu P_Y\nu_i) + g_S^{XYij}({\bar c} P_X b)({\bar\ell_j} P_Y\nu_i)\right\}
\right. \nonumber\\
&&\hspace{4truecm}\left.~+~\sum\limits_X^{L,R}g_T^{XXij}({\bar c}\sigma_{\mu\nu} P_X b)({\bar\ell_j}\sigma_{\mu\nu} P_X\nu_i)\right]\, ,
\label{Hef}
\eea
where $g_{Z}^{XYij}$ represent effective NP Wilson Coefficients with $Z = S,V,T$ referring to Scalar, Vector, or Tensor, $X,Y = L,R$ referring to left-handed and right-handed fermion currents, and $i,j = 1,2,3$ referring to lepton family number (generation). Here, $P_{L,R}$ are projection operators defined as $P_{L,R}=(1\mp \gamma_5)/2$. Note that, unlike in the scalar ($S$) and vector ($V$) operators, the quark and lepton currents in the tensor ($T$) operators necessarily have the same chirality. In what follows, we only consider NP in $\tau$. Therefore, we suppress the generation indices by rewriting $g_{Z}^{XY33}$ as $g_Z^{XY}$.

In our non-minimal SU(5) model, $\mathcal{H}_{\text{eff}}$ receives contributions from tree-level diagrams mediated by the exchange of the $R_{2}$ leptoquark. The most general leptoquark couplings to SM fermions have been discussed in the literature; see, for example, \cite{Sakaki:2013bfa}. The Lagrangian that describes the interactions of $R_{2}$ with SM fermions in our SU(5) model is, 
\bea 
\mathcal{L}_{R_{2}}
  &=&  (Y_2^{UL})_{ij} \ts
    \bar{\hat{u}}_{ a  i}\tts
    P_L \hat{\nu}_{j} R_{2}^{\tts a  2}
   + (Y_2^{EQ})_{ij}(V_{\text{CKM}})_{jk} \ts
    \bar{\hat{e}}_{i}\tts  \ts P_L \hat{d}^{\tts a }_{k} \ts R_{2 a  2}^{\tts*} \hc
\eea
Once the \(R_2\) leptoquark is integrated out, we obtain the following contributions to ${\cal H}_\eff$.
\bea
{\mathcal H}_{\eff}&=&-\frac{1}{2 m^2_{R_{2}}}
\sum^3_{i=1}(Y_2^{UL})_{23} (Y_2^{EQ})_{ 3i}
(V_{\text{CKM}})_{i3}\bigg[(\bar{\tau} P_L \nu_\tau)(\bar{c} P_L
b) \nonumber \\
&&\hspace{7truecm} \left.+~\frac{1}{4}(\bar{\tau} \sigma_{\mu \nu} P_L \nu_\tau)(\bar{c}
\sigma^{\mu \nu} P_L b) \right]\,.
\label{eq:LT21} 
\eea
Comparing this with Eq.~(\ref{Hef}) at the Leptoquark mass scale, $\mu = m_{R_2}$, we find,
\begin{eqnarray}
\label{eq:R2lepto}
  g_S^{LL}=\frac{1}{4 \sqrt{2}\, G_F V _{cb}\,m^2_{R_{2}}}  \sum^3_{i=1}(Y_2^{UL})_{ 2 3 } (Y_2^{EQ})_{ 3i}
(V_{\text{CKM}})_{i3},\,\,\,\,\,\,\,\,\,\,\,\,\,\,\,\,\,\,
g_{T}^{LL}=\frac{1}{4} g_S^{LL},\label{leptq}
\end{eqnarray}
Note that all coefficients except \(g_S^{LL}\) and \(g_T^{LL}\) vanish in this model. We take these as the respective Wilson coefficients also at the bottom quark mass scale, $\mu = m_b$.
Following Refs.~\cite{Iguro:2018vqb, Asadi:2018wea, Asadi:2018sym, Iguro:2024hyk}, with NP restricted to the third generation leptons ($\tau,\nu_\tau$), one can show that the ratios $R_{D^{(*)}}$ take the following forms.
\begin{eqnarray}
R_D &=& R_D^{\rm SM} \Big[1 + 1.01 |g_{S}^{LL}|^2 +0.84
|g_{T}^{LL}|^2 + 1.49 \ \text{Re}( g_{S}^{LL})
 +1.08 \text{Re}( g_{T}^{LL})\Big], \\
R_{D^*} &=&   R_{D^*}^{\rm SM} \Big[1 + 0.04  |g_{S}^{LL}|^2
+16.0 |g_{T}^{LL}|^2- 0.11 \ \text{Re}( g_{S}^{LL}) -5.17
\text{Re}( g_{T}^{LL}) \Big].
\end{eqnarray}
The assumption used to arrive at the above expressions, that NP only affects the third-generation leptons, is motivated by the absence of deviations from the SM in modes involving the light leptons (\( \ell = e, \mu \)). The SM predictions for $R_{D^{(*)}}$ are \cite{Iguro:2024hyk}
\begin{align}
\begin{aligned}
   R_{D}^\textrm{SM} &= 0.290 \pm 0.003\,,\\[0.5em]
   R_{D^\ast}^\textrm{SM}& = 0.248 \pm 0.001.
\end{aligned}
\end{align}
In contrast, the experimental world averages of \( R_{D^{(\ast)}} \) are given by HFLAV \cite{HFLAV:RDRDst2024update} as
\begin{align}
    \begin{aligned}
    R_D &= 0.342 \pm 0.026\,, \\
    R_{D^\ast} & = 0.287 \pm 0.012\,.
    \label{eq:WAvalues}
    \end{aligned}
\end{align}

Tree-level leptoquark exchange also modifies the branching ratio ($\cal B$) of the decay \( B_c^- \to \tau^- \bar{\nu}_{\tau} \), as follows~\cite{Li:2016vvp,Alonso:2016oyd,Iguro:2018vqb,Asadi:2018wea,Asadi:2018sym},
\begin{equation}
{\cal B}(B_c^- \to \tau^- \bar{\nu}_{\tau}) ~=~ {\cal B}(B_c^- \to
\tau^- \bar{\nu}_{\tau})_{\text{SM}}\,\left| 1
 - \frac{m_{B_c}^2 g_{S}^{LL}}{m_\tau(m_b+m_c)} \right|^2 ,
\label{BRBc_TypeII}
\end{equation}
where $m_{B_c}^2/m_\tau(m_b+m_c) = 4.065$ and
\begin{equation}
{\cal B}(B^-_c \to  \tau^-
\bar{\nu}_\tau)_{\rm SM} ~=~ \tau_{B_c} \dfrac{G_F^2}{8\pi}
|V_{cb}|^2 f_{B_c}^2 m_{B_c} m_{\tau}^2  \left(1-
\frac{m_{\tau}^2}{m_{B_c}^2}\right)^2, 
\label{SM_leptonic}
\end{equation}
with $\tau_{B_c}$ and $f_{B_c}$ respectively denoting the lifetime of the $B^-_c$ meson and its decay constant.  

Direct constraints on the leptonic \( B_c \) branching ratios are not available from searches at the LHC. However, an estimate based on LEP data provides a strong upper bound of \({\cal B}(B_c^- \to \tau^- \bar{\nu}_\tau) \leq 10\% \) \cite{Akeroyd:2017mhr}. Subsequent studies using the measured \( B_c \) lifetime have suggested that this bound could be considerably relaxed to \( \leq 39\% \) \cite{Bardhan:2019ljo} and \( \leq 60\% \)~\cite{Blanke:2018yud, Blanke:2019qrx}. In our analysis, we adopt the upper bound \({\cal B}(B_c^- \to \tau^- \bar{\nu}_\tau)_{\rm UB} = 60\% \). Using  this upper bound and $\mathcal{B}(B_c\to \tau\overline\nu)_{\rm SM} \simeq 0.022$, one finds \cite{Blanke:2018yud},
\begin{align}
 \label{eq:Bc}
 \left| 1  - 4.35\, g^{LL}_S \right|^2 = \frac{\mathcal{B}(B_c\to \tau\overline\nu)}{\mathcal{B}(B_c\to \tau\overline\nu)_{\rm SM}} <
27.1\left( \frac{\mathcal{B}(B_c\to \tau\overline\nu)_{\rm UB}}{0.6}\right)\,.
\end{align}

The longitudinal polarization fractions of the $D^*$ in $\bar{B}\to D^*\tau^-\overline{\nu}$, denoted by $F_L^{D^*}$, and the $\tau$ in $\bar{B}\to D^{(*)}\tau^-\overline{\nu}$, given by $P_\tau^{D^{(\ast)}}$, are additional observables that receive contributions from the effective Hamiltonian, \( \mathcal{H}_{\eff} \), of Eq.~(\ref{Hef}). Since some of these observables have been measured, they can provide crucial complementary constraints on the relevant WCs in our model. While their explicit definitions can be found in Refs.~\cite{Tanaka:2012nw, Asadi:2018sym, Iguro:2018vqb}, with non-vanishing WCs \( g_{S}^{LL} \) and \( g_{T}^{LL} \), we find \cite{Iguro:2018vqb, Asadi:2018wea, Asadi:2018sym, Iguro:2024hyk},
\begin{eqnarray}
F_L^{D^*} &=& \frac{F^{D^*}_{L,\,{\rm SM}}}{r_{D^*}} \left[1 + 0.08\,\left|g_{S}^{LL}\right|^2 + 6.90\,\left|g_{T}^{LL}\right|^2- 0.25\,{\rm Re}\left(g_{S}^{LL}\right) - 4.30\, {\rm Re}\left(g_{T}^{LL}\right)\right]\,, \label{FLD_RPV} \\
P_\tau^D &=& \frac{P^D_{\tau,\,{\rm SM}}}{r_{D}} \left[1 + 3.04\,\left|g_{S}^{LL}\right|^2 + 0.17\,\left|g_{T}^{LL}\right|^2 + 4.50\,{\rm Re}\left(g_{S}^{LL}\right) - 1.09\,{\rm Re}\left(g_{T}^{LL}\right)\right]\,, \\
P_\tau^{D^*} &=&  \frac{P^{D^*}_{\tau,\,{\rm SM}}}{r_{D^*}} \left[1 -
0.07\,\left|g_{S}^{LL}\right|^2 - 1.85\,\left|g_{T}^{LL}\right|^2  + 0.23\,
{\rm Re}\left(g_{S}^{LL}\right) - 3.47\,{\rm Re}\left(g_{T}^{LL}\right)\right]\,,
\label{PTAU_RPV}
\end{eqnarray}
where $r_{D^{*}} = R_{D^{(*)}} / R^{\rm
SM}_{D^{(*)}}$. The SM predictions for the above observables are \cite{Iguro:2024hyk},
\bea
F_{L,\,\textrm{SM}}^{D^{\ast}} &=& 0.464 \pm 0.003\,, \\
P_{\tau,\,{\rm SM}}^{D} &=& 0.331 \pm 0.004\,, \\
P_{\tau,\,{\rm SM}}^{D^*} &=& -0.497 \pm 0.007\,.
\eea
In our analysis, we use \( F^{D^*}_{L,\,{\rm Expt}} = 0.49 \pm 0.05 \) \cite{Iguro:2024hyk}, which is the average of the measured values \( F^{D^*}_{L,\,{\rm Belle}} = 0.60 \pm 0.08_{\rm stat} \pm 0.04_{\rm syst}\) \cite{Belle:2019ewo} and \( F^{D^*}_{L,\,{\rm LHCb}} = 0.43 \pm 0.06_{\rm stat} \pm 0.03_{\rm syst} \) \cite{LHCb:2023ssl}. Note that the latter is obtained from a combined LHCb dataset that includes Run 1 and part of Run 2. For the \( \tau \) longitudinal polarization, experimental measurements of \( P^D_\tau \) are not available to date. However, we do have \( P^{D^*}_{\tau,\,{\rm Expt}} = -0.38^{+0.53}_{-0.55} \) \cite{Hirose:2017dxl, Hirose:2016wfn, Adamczyk:2019wyt}. It is important to note that the large experimental uncertainties in \( P^D_\tau \) result in weak constraints on the relevant WCs in our analysis.

The baryonic decay mode \(\Lambda_b \to \Lambda_c \tau \overline{\nu}\) is also governed by the transition \(b \to c \tau \overline{\nu}\) and, therefore, receives new contributions from the effective Hamiltonian, \({\mathcal H}_{\text{eff}}\), given in Eq. (\ref{Hef}). This decay can also be a sensitive probe of new physics \cite{Shivashankara:2015cta} and we investigate the size of the leptoquark contributions to the ratio \(R_{\Lambda_c}\), defined as
\beq
R_{\Lambda_c} ~\equiv~ \frac{\mathcal{B}(\Lambda_b \to \Lambda_c \tau \overline{\nu})}{\mathcal{B}(\Lambda_b \to \Lambda_c \ell \overline{\nu})}\,.
\eeq%
An expression for \(R_{\Lambda_c}\), incorporating lattice QCD results for the transition \(\Lambda_b \to \Lambda_c\) \cite{Detmold:2015aaa, Datta:2017aue, Murgui:2019czp} as well as numerical values of other measurable quantities, is provided in Refs. \cite{Fedele:2022iib, Iguro:2024hyk}. Keeping only terms that depend on $g_S^{LL}$ and $g_T^{LL}$, the only relevant nonzero WCs in our model, we find \cite{Fedele:2022iib,Iguro:2024hyk,Duan:2024ayo},
\beq 
R^{R_{2}}_{\Lambda_c} ~=~ R_{\Lambda_c}^{\rm
SM}\left[1 + 0.32\,\left|g^{LL}_{S}\right|^2  + 10.4\,\left|g^{LL}_T\right|^2  + 0.33 \,{\rm Re}\left(  g^{LL*}_{S}\right)  
- 3.11\,{\rm Re}\left(g^{LL*}_T\right)\right] \,, 
\eeq%
where $R_{\Lambda_c}^{\rm SM} = 0.324 \pm 0.004$ accounts for the SM prediction 
\cite{Bernlochner:2018kxh,Bernlochner:2018bfn}. The LHCb collaboration has reported the observed value of the ratio as
$R_{\Lambda_c}^{\rm LHCb} = 0.242 \pm  0.026_{\rm stat} \pm 0.071_{\rm syst}$
\cite{LHCb:2022piu}. As discussed in Ref.\cite{Iguro:2024hyk}, normalizing with the SM prediction of $\Gamma(\Lambda_b\to\Lambda_c\mu\overline{\nu})$ refines the accuracy and marginally raises the central value,
$R_{\Lambda_c} = |0.041/V_{cb}|^2 (0.271 \pm 0.069) = 0.271 \pm 0.072$ \cite{Bernlochner:2022hyz}. 

\begin{figure}[!htbp]
\begin{center}
\includegraphics[width=0.7\textwidth]{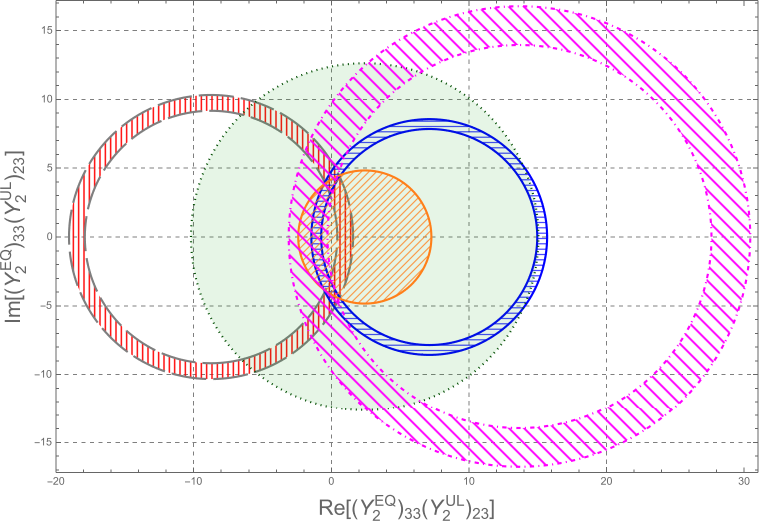} \vskip10pt
\includegraphics[width=0.4\textwidth]{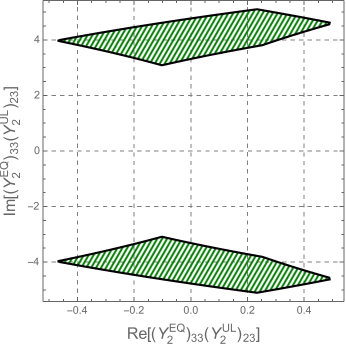} \hspace{5truemm}
\includegraphics[width=0.4\textwidth]{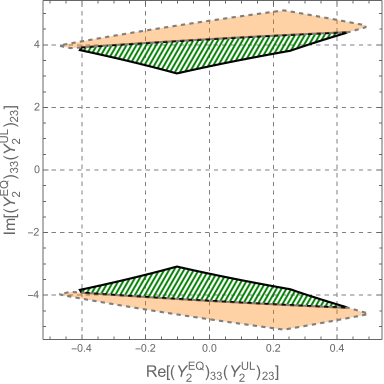}
\end{center}
\caption{Allowed parameter space spanned by the real and imaginary parts of the product $(Y_2^{EQ})_{33}(Y_2^{UL})_{23}$ for an $R_2$ leptoquark with mass $m_{R_2} = 2$ TeV. The top figure shows the allowed region based on various measurements: the upper bound on ${\cal B}( B_c \to \tau \overline{\nu})$ (filled green, dotted boundary), $R_{D}$ (red, gray dashed boundary, vertical hatching), $R_{D^{*}}$ (blue, horizontal hatching), $F_{L}(D^{*})$ (magenta, dot-dashed boundary, diagonal hatching), and $R_{\Lambda_{c}}$ (orange, diagonal hatching). We use a 90\% confidence-level (c.l.) bound for $R_{\Lambda_c}^{\rm LHCb}$ and respective 1$\sigma$ bounds for the other observables. The bottom-left figure shows the allowed region (green, hatched) of parameter space that satisfies several bounds -- the upper bound on ${\cal B}(B_c\to\tau\overline{\nu})$, $R_{D}$, $R_{D^{*}}$, $F_{L}(D^{*})$, and $P_{\tau}(D^{*})$. The bottom-right figure also shows the region excluded by the 90\% c.l.~bound on $R_{\Lambda_c}^{\rm LHCb}$ (orange, dashed boundary).
\label{fig:parspace}}
\end{figure}

The several observables discussed thus far can now be simultaneously used to constrain the WCs $g_S^{LL}$ and $g_T^{LL}$. Since these WCs can be expressed in terms of the Yukawa couplings of the $R_2$ leptoquark, as shown in Eq.~(\ref{eq:R2lepto}), one can in turn impose constraints on the Yukawa couplings. Here we note that the dominant contribution to the WCs comes from the term proportional to $(V_{\rm CKM})_{33} = V_{tb} = 1$. Thus, ignoring the CKM suppressed terms in Eq.~(\ref{eq:R2lepto}) and using $1\,\sigma$ bounds for the observables discussed above (90\% confidence level or 1.65$\sigma$ bound for $R_{\Lambda_c}^{\rm LHCb}$), we obtain an allowed region of parameter space spanned by the real and imaginary parts of the product $(Y_2^{EQ})_{33}(Y_2^{UL})_{23}$. We present our results in Figure \ref{fig:parspace}. 
 
We also perform a fit to the data (excluding $R_{\Lambda_c}$) to constrain the complex-valued product $(Y_2^{UL})_{ 2 3 } (Y_2^{EQ})_{ 33}$ and find the following best-fit values for its real and imaginary parts,
\bea
{\rm Re}[g_{S}^{LL}]=0.012\pm 0.032\,\qquad {\rm Re}\left[(Y^{UL}_2)_{23}(Y^{EQ}_2)_{33}\right] &=& 0.13 \pm 0.33 \,, \label{eq:bfvr} \\
{\rm Im}[g_{S}^{LL}]=0.46\pm 0.03\,\qquad {\rm Im}\left[(Y^{UL}_2)_{23}(Y^{EQ}_2)_{33}\right] &=& 4.81 \pm 0.31 \,, \label{eq:bfvi}
\eea
with $\chi^2_{\rm min}/{\rm d.o.f} = 1.3$ for a $p$-value of 0.24. This estimate of the Yukawa couplings uses $m_{R_2} = 2$ TeV and the $\vert V_{cb} \vert_{\text{exclusive}} $ value. In our analysis, we take the product of the Yukawa couplings to have an upper bound of $\sim 4\pi$ \footnote{In perturbative quantum field theory, loop corrections are generally proportional to the square of the coupling divided by $4\pi$. Thus, the upper bound on the product of couplings is typically set to $4\pi$ to preserve perturbativity.} One could also adopt a more conservative upper bound on the Yukawas $\sim 1$ in which case our model can partially explain the present experimental data.
We note that the above determination of the Yukawa coupling uses low-energy observables where the effective theory defined at $m_{b}$-scale in Eq.~\eqref{eq:LT21} and Eq.~\eqref{eq:R2lepto} are applicable. However, limits on Yukawa couplings inferred in an effective theory framework from direct searches at LHC cannot be readily applied for leptoquark masses $m_{R_{2}} < 10 \,\text{TeV}$ as shown in Ref~\cite{Iguro:2020keo,Endo:2021lhi} and therefore those limits are not relevant in our case with the leptoquark mass, $m_{R_2} = 2$ TeV. It is also observed that explicit leptoquark models with lower leptoquark masses result in weaker sensitivity to the Yukawa couplings from nonresonant searches at LHC~\cite{Marzocca:2020ueu,Endo:2021lhi}. In our model analysis, we avoid collider constraints from the LHC in $pp\to \tau\tau, \tau\nu$ channels by setting $(Y_2^{UL})_{13} = 0$. Since $(Y_{2}^{UL})_{23}$ is required to be non-vanishing to explain $b\to c \ell \bar{\nu}$ data, it also contributes to $Z\to \tau^{+} \tau^{-}$ decays at 1-loop. The contribution to the branching fraction of $Z\to \tau^{+}\tau^{-}$ from $(Y_{2}^{UL})_{23}$ is discussed in Appendix \ref{app:ztautauconstrains}. There we conclude that the measured value of the $Z\to \tau^{+}\tau^{-}$ branching fraction is consistent with the allowed parameter space from $b\to c \ell \bar{\nu}$ data (see Fig.~\eqref{fig:ztautau}).

Using these best-fit values, we find
\beq
R_{\Lambda_{c}} ~=~ 0.391\pm 0.007. \label{eq:rlc}
\eeq%
Comparing the above value of $R_{\Lambda_c}$ with its measured value from LHCb, $R_{\Lambda_{c}} = 0.242\pm 0.026 \pm 0.071$ \cite{LHCb:2022piu}, one notices that the central value of the measurement is almost $1.5\,\sigma$ smaller than that obtained in Eq.~(\ref{eq:rlc}). This aligns well with the observation that the suppression in the experimental value of $R_{\Lambda_{c}}$ is in tension with the NP scenarios preferred by the $R_{D^{(*)}}$ anomaly. However, a large experimental uncertainty in $R_{\Lambda_{c}}$ prevents us from reaching a firm conclusion in this case.

One final point deserves further attention. In addition to charged-current transitions, the $R_2$ leptoquark can also mediate neutral-current $B$-decays at tree level. In particular, the neutral-current transition $b \to s \ell^-_i \ell^+_j$ will receive contributions from the following interaction Lagrangian.
\bea \mathcal{L}_{R_{2}}
  &=& - (Y_2^{EQ})_{ij}  (V_{\text{CKM}})_{j}{}^{k}\ts
    \bar{\hat{e}}_{i}\tts \ts P_L \hat{d}^{\tts a }_{k} \ts R_{2 a  2}^{\tts *}-(Y_2^{EQ})^*_{ij}  (V_{\text{CKM}})_{j}{}^{k*}\ts
    \bar{\hat{d}}^{\tts a }_{k}\tts \ts P_R  \hat{e}_{i} \ts R_{2 a  2}^{\tts }\,. \eea
Once the leptoquark, $R_2$, is integrated out, the effective Hamiltonian takes the form,
\beq
{\cal H}_{\rm eff} ~=~ \frac{1}{2m^2_{ R_{2}}} \sum\limits^3_{k,\ell=1} (Y_2^{EQ})_{ik}(V_{\rm CKM})_{k3}
(Y_2^{EQ})^*_{j\ell}(V^*_{\rm CKM})_{\ell 2} (\bar{s}\gamma^\mu
P_L b) (\bar{\ell}_i\gamma^\mu P_R \ell_j)\,.
\eeq%
Comparing this with the standard dimension-6 effective Hamiltonian for $b\to s\ell^+_i\ell^-_j$ \cite{Buchalla:1995vs}, one finds the relevant WCs to be,
\beq
C_9^{ij} ~=~ C_{10}^{ij} ~=~ -\frac{\pi}{\sqrt{2}\,G_F\,V_{tb}\,V^*_{ts}\,\alpha\, m^2_{R_{2}}} \sum\limits_{k,\ell = 1}^3 (Y_2^{EQ})_{ik} (V_{\rm CKM})_{k3} (Y_2^{EQ})^*_{j\ell}(V^*_{\rm CKM})_{\ell 2}\,.
\eeq%
Once again, several Yukawa couplings are involved in the above relations. We can use this to our advantage. For example, assuming the $Y_2^{EQ}$ couplings to be vanishingly small for all three lepton flavors, one can avoid effects in charged-current $B$-anomalies, suppress contributions to \( b \to s \ell^+_i \ell^-_j \), as well as avoid bounds on the $R_2$ leptoquark mass from LHC direct searches for leptoquark decays to a quark and a light charged lepton \cite{ATLAS:2020dsk}.

\section{$B^+\to K^+\nu{\bar\nu}$ in non-minimal SU(5)}

In this section, we compute the contribution from the non-minimal SU(5) model to $b\to s\nu\bar{\nu}$. The SM prediction for the branching fraction of \( B^+ \to K^+ \nu \nu \) is \cite{Parrott:2022zte}:
\beq
\mathcal{B}_{\text{SM}}(B^+ \to K^+ \nu \nu) = (5.58 \pm 0.37) \times 10^{-6}\,.
\label{SMpred}
\eeq%
Recently, the Belle II collaboration reported using the novel Inclusive Tagging Analysis (ITA) method to measure \cite{Belle-II:2023esi}:
\beq
\mathcal{B}(B^+ \to K^+ \nu \nu) = (2.7 \pm 0.5 \, (\text{stat}) \pm 0.5 \, (\text{syst})) \times 10^{-5}. \label{eq:b2kinvobs}
\eeq
The measured value of \({\cal B}(B^+\to K^+\nu{\bar\nu})\) exceeds the SM prediction by approximately \( 2.7\sigma\). This observed excess could be due to NP, and as such it has generated significant theoretical interest in potential NP explanations for the discrepancy \cite{Athron:2023hmz,Bause:2023mfe,Allwicher:2023xba,Felkl:2023ayn,Dreiner:2023cms,Abdughani:2023dlr,He:2023bnk,Berezhnoy:2023rxx,Datta:2023iln,Altmannshofer:2023hkn,McKeen:2023uzo,Fridell:2023ssf,Ho:2024cwk,Gabrielli:2024wys,Li:2024thq,Hou:2024vyw,He:2024iju,Bolton:2024egx,Marzocca:2024hua,Aghaie:2024jkj,Rosauro-Alcaraz:2024mvx,Eguren:2024oov,Buras:2024ewl,Hati:2024ppg,Wang:2024prt}.

\begin{figure}[!htbp]
\begin{center}
\includegraphics[width=0.48\textwidth]{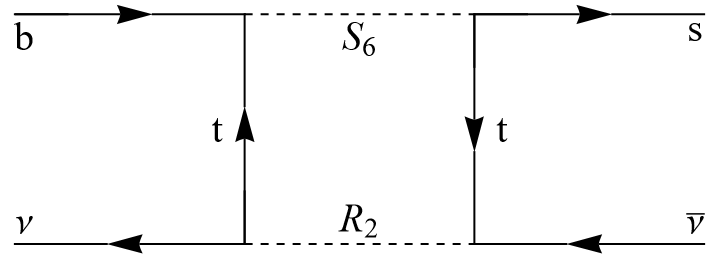}
\includegraphics[width=0.48\textwidth]{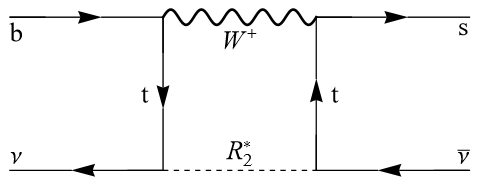} \vskip 10pt
\includegraphics[width=0.48\textwidth]{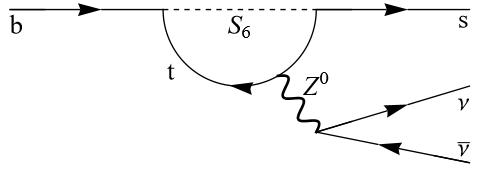}
\end{center}
\caption{One-loop Feynman diagram contributions to $b\to s \nu{\bar\nu}$. The top-left diagram is a box diagram generated via $S_6$ and $R_2$ exchanges. The top-right diagram is another box diagram that incorporates $W^+$ and $R_2^*$ exchanges. The bottom diagram is a $Z^0$-penguin diagram with the $S_6$ in the loop.
\label{box}}
\end{figure}

In our non-minimal SU(5) model with a \(\mathbf{45}\)-dimensional Higgs, contributions to this transition arise from an $S_6$--$R_2$ box diagram, a $W$--$R_2$ box diagram, and a $Z^0$-penguin diagram, as illustrated in Fig.~\ref{box}.
These diagrams depend on different combinations of Yukawa couplings, so that comparisons among them are generally not possible. However, ignoring all constraints and considering all Yukawas to be of order unity, we find the following hierarchy among these diagrams: the $Z^0$-penguin gives the largest contribution, followed by the $S_6$--$R_2$ box diagram, while the $W$--$R_2$ box diagram generates the smallest contribution of the three.

Once the heavy particles in the diagrams of Fig.~\ref{box} -- the diquark, leptoquark, $W^+$, $Z^0$, and top quark -- are integrated out, the following effective Hamiltonian governing the $b\to s\nu{\bar\nu}$ transition \cite{Buras:2014fpa} emerges.
\begin{equation} \label{eq:Heff1}
{\mathcal{H}}_{\text{eff}} = -
\frac{4\,G_F}{\sqrt{2}}V_{tb}V_{ts}^*\left(C_L \mathcal O_L +C_R
\mathcal O_R  \right) ~+~ \text{h.c.} \,,
\end{equation}
where the operators ${\cal O}_{L,R}$ are given as,
\begin{align}
\mathcal{O}_{L} &=\frac{e^2}{16\pi^2} (\bar{s}  \gamma_{\mu} P_L
b)(  \bar{\nu} \gamma^\mu(1- \gamma_5) \nu) \,,& \mathcal{O}_{R}
&=\frac{e^2}{16\pi^2}(\bar{s}  \gamma_{\mu} P_R b)(  \bar{\nu}
\gamma^\mu(1- \gamma_5) \nu) \,.
\end{align}
The WCs, \( C_{L,R} \), in the above effective Hamiltonian can be expressed as \( C_L = C_L^{\rm SM} + C_L^{\rm NP} \) and \( C_R = C_R^{\rm NP} \). The SM contribution to the WC, \( C_L^{\rm SM} \), is well known, with high precision, including next-to-leading order QCD corrections \cite{Buchalla:1993bv,Misiak:1999yg,Buchalla:1998ba} and two-loop electroweak contributions \cite{Brod:2010hi}. It can be expressed as \( C_L^{\text{SM}} = -X_t/s_w^2 \) with \( X_t = 1.469 \pm 0.017 \) and \( s_w^2 = 0.23121 \), so that \( C_L^{\text{SM}} \approx -6.35\). The non-minimal SU(5) model introduces new contributions to the WCs \( C_L^{\text{NP}} \) and \( C_R^{\text{NP}} \). 
In calculating the amplitude for \( B^+\to K^+ \nu{\bar\nu} \) using the effective Hamiltonian, one notices that the hadronic matrix elements \( \langle K | \bar{s} \gamma^\mu P_{L} b | B \rangle \) and \( \langle K | \bar{s} \gamma^\mu P_{R} b | B \rangle \) are equal, since the axial-vector current, \(\langle K|\bar{s} \gamma^\mu\gamma^5 b|B\rangle\), vanishes. Therefore, the total amplitude for \(B^+\to K^+\nu{\bar\nu}\) is proportional to \(C_L + C_R\) and consequently its branching ratio is proportional to \(|C_L + C_R|^2\). In the SM, this factor is \(|C_L^{\rm SM}|^2\), while in our non-minimal SU(5) model it is \(|C^{\rm SM}_L + C^{\rm NP}_L + C^{\rm NP}_R|^2\). Thus, one can express the NP-enhanced branching ratio for \(B^+\to K^+\nu{\bar\nu}\) as,
\bea
\label{eq:CLCRratio}
{\cal B}_{\rm NP}(B^+\to K^+\nu{\bar\nu}) &=& {\cal B}_{\rm SM}(B^+\to K^+\nu{\bar\nu}) \left|1+\frac{C^{\rm NP}_L + C^{\rm NP}_R}{C^{\rm SM}_{L}}\right|^2\,.
\eea

The contributions from the $S_6$--$R_2$ box diagram shown in the top-left panel of Fig.~\ref{box} are,
\bea 
(C^{NP}_{L})_{S_{6}-R_{2}}&\approx & - \frac{ V_{cb}}{32\pi\,\alpha\,g^{2} z\,V_{tb}V_{ts}^*}\,f(x_{S_6},x_{R_2})\,(Y^{QQ}_6)_{32}  (Y^{QQ}_6)^*_{32}\sum\limits^{3}_{i=1} |(Y^{UL}_2)_{3i}|^2\,, \label{eq:CLbox} \\
(C^{NP}_{R})_{S_{6}-R_{2}}&\approx & \frac{1}{16\pi\,\alpha g^{2} z\,V_{tb}V_{ts}^*}\,g(  x_{S_6},x_{R_2})\,(Y^{DU}_6)^*_{33} (Y^{DU}_6)_{23}\sum\limits^{3}_{i=1}|(Y^{UL}_2)_{3i}|^2\,, \label{eq:CRbox}
\eea
where $x_{S_6(R_2)}=m^2_t/m^2_{S_6(R_2)}$, $z=m^{2}_{t}/m_{W}^{2}$, $g$ is the weak coupling, and $f(x_{S_6},x_{R_2})$ and $g(x_{S_6},x_{R_2})$ are loop functions that can be expressed as,
\bea
\label{eq:pureloop}
f(x,y) &=& -\frac{x\,y}{x-y} \left[\frac{1-x + x\ln x}{(1-x)^2} - x\to y\right]\,,\\
g(x,y) &=& -  \frac{x\,y}{x-y} \left[\frac{1-x +\ln x}{(1-x)^2} ~-~ x\to y\right]\,.
\eea
Note that the loop functions, \(f(x,y)\) and \(g(x,y)\) are symmetric under the interchange of \(x\) and \(y\). In the limiting case obtained when the diquark and leptoquark have the same mass, the loop functions take the forms,
\beq
f(x,x) = -\frac{x^2[2(1-x)+(1+x)\ln x]}{(1-x)^3}\,,~~ g(x,x) = -\frac{x(1-x^2+2x\ln x)}{(1-x)^3}\,.
\eeq
To simplify our analysis, we have retained only the leading contribution to $C_{L}^{NP}$ from products of $Y^{QQ}_6$ components and the relevant CKM matrix elements, as given by Eq.~\eqref {eq:LYbelowGUT}. 

The WC contributions from the $W$--$R_2$ box diagram shown in the top-right panel of Fig.~\ref{box} are as follows.
\beq
(C^{NP}_{L})_{W-R_{2}} = -\frac{1}{4\pi\alpha\,z}\sum\limits_{i=1}^{3}\left|(Y_{2}^{UL})_{3i}\right|^{2}f(z,x_{R_{2}})\,,~~
(C^{NP}_{R})_{W-R_{2}} = 0\,, \label{eq:CWbox}
\eeq
where the loop function $f(x,y)$ is defined in Eq.~\eqref{eq:pureloop}. Finally, the WC contributions from the $Z^0$-penguin diagram shown in the bottom panel of Fig.~\ref{box} can be evaluated using $c_R = -2s^2_w/3 \approx -0.154$ and $c_L = 1/2 + c_R \approx 0.346$, as
\bea
(C^{NP}_{L})_{Z-S_{6}}&=& \frac{V_{cb}}{2\pi \alpha \, V_{tb}^{}V_{ts}^*}(Y^{QQ}_6)_{32}(Y^{DU}_6)^*_{23}\left(c_L \tilde f(x_{S_6}) + c_R \tilde g(x_{S_6})\right) \,, \label{eq:CLp}\\
(C^{NP}_{R})_{Z-S_{6}}&=& \frac{1}{2\pi \alpha V_{tb}^{}V_{ts}^*}(Y^{DU}_6)_{33}(Y^{QQ}_6)^*_{32}\left(c_R \tilde f(x_{S_6}) + c_L \tilde g(x_{S_6})\right) \label{eq:CRp}\,,
\eea
where once again the loop functions $\tilde f(x)$ and $\tilde g(x)$ can be expressed as,
\beq
\tilde f(x) = -\frac{x (1-x+\ln{x})}{(1-x)^2}\,,~~
\tilde g(x) = -\frac{1-x^{2}+2\ln{x}}{4(1-x)^{2}}-\frac{1}{2}\ln{\left(\frac{\mu^2}{m^2_t}\right)}\,.
\eeq

Once the above expressions are collected, the total contribution to $b\to s\nu\bar{\nu}$ can be obtained by summing over the three diagrams in Fig.~\ref{box}. The total depends on various combinations of coupling constants, some of which are constrained from experiments while some remain free. Since there are many possibilities, we simplify our discussion by first listing the constraints on the Yukawa couplings that appear in the various diagrams:
\begin{enumerate}
    \item In a model with significantly large $(Y_2^{UL})_{33}$, one can obtain stringent constraints on the mass of the $R_2$ leptoquark from LHC direct searches for a single top with missing energy, i.e., the top quark and neutrino channel \cite{CMS:2018qqq}. Thus, in our model, we set $(Y_2^{UL})_{33} = 0$ to avoid such direct constraints on the leptoquark mass.

    \item  Considering the $Z^0$-penguin diagram, the bound from \( b \to s\gamma \) puts a constraint on the following product of Yukawa couplings: \( |(Y^{DU}_6)_{33}(Y^{QQ}_6)_{32}| \lesssim 2.7 \times 10^{-3} \) (see the discussion in Section \ref{sec:nonminSU5}). This ensures that $(C^{NP}_{R})_{Z-S_{6}}$ is bounded from above. 

    \item Neither \( B_{s(d)} \) mixing nor \( b \to s(d)\gamma \) impose constraints on the quantity \(|(Y^{UL}_2)_{3i}|^2\) appearing in \( C^{\rm NP}_{L,R}\). Similarly, \(|(Y^{UL}_2)_{3i}|^2\) is not constrained by \( D \) meson mixing, as the relevant constraint applies to the combination \( (Y^{UL}_2)_{2i} (Y^{UL}_2)_{2j} (Y^{UL}_2)^*_{1i} (Y^{UL}_2)^*_{1j} \).  Consequently, \( |(Y^{UL}_2)_{3i}|^2 \) remains a free parameter. Note that this choice is also consistent with our findings from  $b\to c \ell \bar{\nu}$ data (see Fig.~\ref{fig:ztautau} in Appendix \ref{app:ztautauconstrains}).
\end{enumerate}

We now consider the following two scenarios:
\begin{itemize}
\item Scenario I: Consistent with 1 and 3 above, we choose \( (Y^{UL}_2)_{3i}^2 \) to vanish. In this case, neither box diagram contributes. Now, considering the $Z^0$ penguin and using the bound from \( b \to s\gamma \), \( |(Y^{DU}_6)_{33}(Y^{QQ}_6)_{32}| \lesssim 2.7 \times 10^{-3} \), we find, 
\bea
(C^{NP}_{R})_{Z-S_{6}}\lesssim 0.504 \pm 0.011\,.
\eea
Given the largest value of $(C^{NP}_{R})_{Z-S_{6}}$ allowed, we can now put an upper bound on $(C^{NP}_{L})_{Z-S_{6}}$ by saturating the observed $B^{+}\to K^{+}\nu\bar{\nu}$ decay rate in Eq.~\ref{eq:b2kinvobs}. This, in turn, leads to the following upper bound on the product of Yukawa couplings, $(Y^{QQ}_6)_{32}(Y^{DU}_6)^*_{23}$:
\bea
\left|(Y^{QQ}_6)_{32}(Y^{DU}_6)^*_{23}\right| \lesssim 2.8\pm 0.8 \,.
\eea
This estimate is based on $C_{L}^{NP}=(-7.0\pm2.0)$, obtained from Eq.~\eqref{eq:CLCRratio} with $C_{L}^{NP}$ set to have the same sign as $C_{L}^{SM}$. Here, we point out that the $Z^0$ penguin can also contribute to $b\to s\ell^+\ell^-$ processes. Theory predictions of $b\to s\ell^+\ell^-$ are difficult due to the presence of unknown long-distance SM contributions. Requiring no or small contributions to $b\to s\ell^+\ell^-$ will strongly constrain the $Z^0$ penguin contribution to $b\to s\nu{\bar\nu}$.

\item Scenario II: We choose $(Y^{QQ}_6)_{32}=0$ in which case the $Z^0$-penguin vanishes and so does $(C^{NP}_{L})_{S_{6}-R_{2}}$. 
Therefore, the only non-zero WCs in this case are $(C^{NP}_{L})_{W-R_{2}}$ given in Eq.~(\ref{eq:CWbox}) and $(C^{NP}_{R})_{S_{6}-R_{2}}$ given in Eq.~(\ref{eq:CRbox}). Both of these WCs depend on the sum over three leptoquark couplings, which we can set to one. In addition $(C^{NP}_{R})_{S_{6}-R_{2}}$ depends on the product $(Y^{DU}_6)^*_{33}(Y^{DU}_6)^{}_{23}$, which can also be assumed to be of order unity. With these assumptions and setting the diquark and leptoquark masses to 2 TeV, we find, 
\beq
(C^{NP}_{L})_{W-R_{2}} \simeq -0.198\,,~~ (C^{NP}_{R})_{S_{6}-R_{2}} \simeq -4.48\,.
\eeq
This yields $(C^{\rm NP}_L + C^{\rm NP}_R)/C^{\rm SM}_L \sim 74\%$, which in turn implies a factor of three enhancement of \({\cal B}(B^+\to K^+\nu{\bar\nu})\) over its SM value. With these choices made, we find ${\cal B}(B^{+}\to K^{+}\nu\bar{\nu}) \approx 1.68 \times 10^{-5}$, which is comparable to the measured value of this branching ratio given in Eq.~(\ref{eq:b2kinvobs}). Since the WCs depend on products of two or four Yukawa couplings, tiny modifications to multiple Yukawa couplings may lead to a large enhancement in the NP WCs. This, in turn, can significantly enhance the branching ratio.
\end{itemize}

\section{Conclusions \label{sec:conclusions}} 
Understanding the masses and mixing of the quarks and leptons as well as the flavor structure of the SM is one of the key challenges in particle physics. A proposed class of solutions to address this problem is the grand unified theory (GUT) models, of which the SU(5) model is well known. In this work, we considered a non-minimal SU(5) GUT framework, which introduces a 45-dimensional Higgs representation that includes new scalar states such as the leptoquark, \(R_2\), and the diquark \(S_6\). We discussed the effect of these new states in flavor-changing neutral current (FCNC) decays of the $b$ quark. In particular, we focused on the $B$ anomalies, which have generated a lot of interest over the past decade. We showed that the exchange of the diquark at the tree level can lead to new structures in the FCNC effective Hamiltonian for non-leptonic decays. These new contributions can explain the long-standing $\btopiK$ puzzle without conflicting with constraints from other FCNC processes like $B$ mixing and $b\to s\gamma$. In the charged current sector, we demonstrated that the tree-level exchange of the leptoquark, \( R_2 \), can explain the anomalies in the $ \bctaunu$ decays such as the $R_D$ and $R_{D^{*}}$ measurements. Finally, we showed that the loop-level contributions via the $Z^0$-penguin, $W-R_2$  and the $S_6-R_2$ box diagrams, can contribute to the $b\to s\nu{\bar\nu}$ processes and can potentially explain the observed anomaly in the rate of $B^+\to K^+ + {\rm inv}$.

It is worth noting that within the present non-minimal $SU(5)$ setup, which includes light scalar states such as the sextet, leptoquark, and octet, gauge coupling unification can be achieved at a scale of order $2 \times 10^{15}\,\mathrm{GeV}$, as illustrated in Appendix~C. This scale lies somewhat below the value typically required to adequately suppress proton decay mediated by heavy gauge bosons. Consequently, this suggests that additional ingredients---most notably supersymmetry---may be necessary to raise the unification scale to a phenomenologically acceptable range while maintaining consistency with proton decay constraints.

\section*{Acknowledgments}
This work was financially supported by the U.S. National Science Foundation under Grant No.\ PHY-2310627 (BB) and PHY-2309937 (AD). The work of S.~K.~is partially supported by the Science, Technology $\&$ Innovation Funding Authority (STDF) under grant number 48173.

\appendix
\section{Inputs used for $B\to K\pi$ NP fits}
\label{app:inputs}

The non-zero hadronic matrix elements important to express the $B\to K\pi$ decay amplitudes are given by,
\begin{align}
    \langle \pi^{+}(p_{\pi^{+}},m_{\pi^{+}}) \vert \bar{u}\gamma^{\mu}b\vert \bar{B}^{0}(p_{\bar{B}^{0}},m_{\bar{B}^{0}})\rangle &=F_{+}^{\bar{B}^{0}\to\pi^{+}}(q^{2})\big[(p_{\bar{B}^{0}}+p_{\pi^{+}})^{\mu}-\frac{(m_{B^{0}}^{2}-m_{\pi^{+}}^{2})}{q^{2}}q^{\mu}\big]
    \\& \qquad + F_{0}^{\bar{B}^{0}\to\pi^{+}}(q^{2})\frac{(m_{B^{0}}^{2}-m_{\pi^{+}}^{2})}{q^{2}}q^{\mu}\nonumber\\
\langle \pi^{+}(p_{\pi^{+}},m_{\pi^{+}}) \vert \bar{u}b\vert \bar{B}^{0}(p_{\bar{B}^{0}},m_{\bar{B}^{0}})\rangle &= F_{0}^{\bar{B}^{0}\to\pi^{+}}(q^{2})\frac{m_{B^{0}}^{2}-m_{\pi^{+}}^{2}}{m_{b}-m_{u}}\\
\langle 0 \vert \bar{s} \gamma^{\mu}\gamma_{5}u\vert K^{+}(p_{K^{+},m_{K^{+}}})\rangle &=i f_{K^{+}}p_{K^{+}}^{\mu}\\
\langle 0 \vert \bar{s} \gamma_{5}u\vert K^{+}(p_{K^{+},m_{K^{+}}})\rangle &=-i f_{K^{+}}\frac{m_{K^{+}}^{2}}{m_{u}+m_{s}}
\end{align}
where $q^{\mu}=(p_{\bar{B}^{0}}-p_{\pi^{+}})^{\mu}$ and 
\begin{align}
   \langle \bar{K}^{0}(p_{\bar{K}^0},m_{K^0}) \vert \bar{s}\gamma^{\mu}b\vert \bar{B}^{0}(p_{\bar{B}^{0}},m_{\bar{B}^{0}})\rangle &=F_{+}^{\bar{B}^{0}\to \bar{K}^{0}}(\tilde{q}^{2})\big[(p_{\bar{B}^{0}}+p_{K^{0}})^{\mu}-\frac{(m_{B^{0}}^{2}-m_{K^0}^{2})}{\tilde{q}^{2}}\tilde{q}^{\mu}\big]
    \\& \qquad + F_{0}^{\bar{B}^{0}\to \bar{K}^0}(\tilde{q}^{2})\frac{(m_{B^{0}}^{2}-m_{K^{0}}^{2})}{\tilde{q}^{2}}\tilde{q}^{\mu}\nonumber\\
\langle \bar{K}^{0}(p_{\bar{K}^0},m_{K^0}) \vert \bar{s}b\vert \bar{B}^{0}(p_{\bar{B}^{0}},m_{\bar{B}^{0}})\rangle &= F_{0}^{\bar{B}^{0}\to \bar{K^{0}}}(\tilde{q}^{2})\frac{m_{B^{0}}^{2}-m_{K^{0}}^{2}}{m_{b}-m_{s}}\\
\langle 0 \vert \bar{u} \gamma^{\mu}\gamma_{5}u\vert \pi^{0}(p_{\pi^{0},m_{\pi^{0}}})\rangle &=\frac{i}{\sqrt{2}} f_{\pi^{0}}p_{\pi^{0}}^{\mu}\\
\langle 0 \vert \bar{u} \gamma_{5}u\vert \pi^{0}(p_{\pi^{0},m_{\pi^{0}}})\rangle &=-i f_{\pi^{0}}\frac{m_{\pi^{0}}^{2}}{2\sqrt{2}m_{u}}
\end{align}
where $\tilde{q}^{\mu}=(p_{\bar{B}^{0}}-p_{\bar{K}^{0}})^{\mu}$. The numerical values of $F_{+}^{\bar{B}^{0}\to\pi^{+}}$ and $F_{+}^{\bar{B}^{0}\to \bar{K}^{0}}$ are taken from Ref~\cite{Ball:2004ye}.

%%%%%%%%%%%%%%%%%%%%%%%%%%%%%%%%%%%%%%%%%%%%%%%%%%
\section{Constraints from the $Z \to \tau \bar{\tau}$ branching Fraction}
\label{app:ztautauconstrains}

The branching fraction of the $Z$ boson into a tau-antitau pair, $\mathcal{B}(Z \to \tau \bar{\tau})$, is given by \cite{Arnan:2019olv,Becirevic:2024pni}
\beq
\mathcal{B}(Z\to \tau \bar{\tau}) = \dfrac{m_Z
\lambda^{1/2}_Z}{6\pi v^2 \Gamma_Z}
\left[\left(|g_{\tau_L}^{\tau\tau}|^2+|g_{\tau_R}^{\tau\tau}|^2\right)\Bigg{(}1-
\dfrac{ m_{\tau}^2}{m_Z^2}\Bigg{)} + 6 \dfrac{m^2_{\tau} }{m_Z^2}\,
\mathrm{Re}\left[g_{\tau_L}^{\tau\tau} \left(
g_{\tau_R}^{\tau\tau}\right)^\ast\right] \right]\,,
\eeq%
where $m_{\tau}$ is the tau mass and $\lambda_Z\equiv m_Z^2
(m_Z^2-4 m_{\tau}^2)$ and
%%%%%%%%%%%%%%
\beq g_{\tau_{L(R)}}^{\tau \tau} =
g_{\tau_{L(R)}}^{\mathrm{SM}}+\delta g_{\tau_{L(R)}}^{\tau
\tau}\,, \eeq%
%%%%%%%%%%%%%%
with $g_{\tau_L}^{\mathrm{SM}} = -\frac{1}{2}+\sin^2\theta_W$ and
$g_{\tau_R}^{\mathrm{SM}}=\sin^2\theta_W$ with $\theta_W$ is the
Weinberg angle. 

\begin{figure}[!htbp]
\includegraphics[width=8cm,height=7cm]{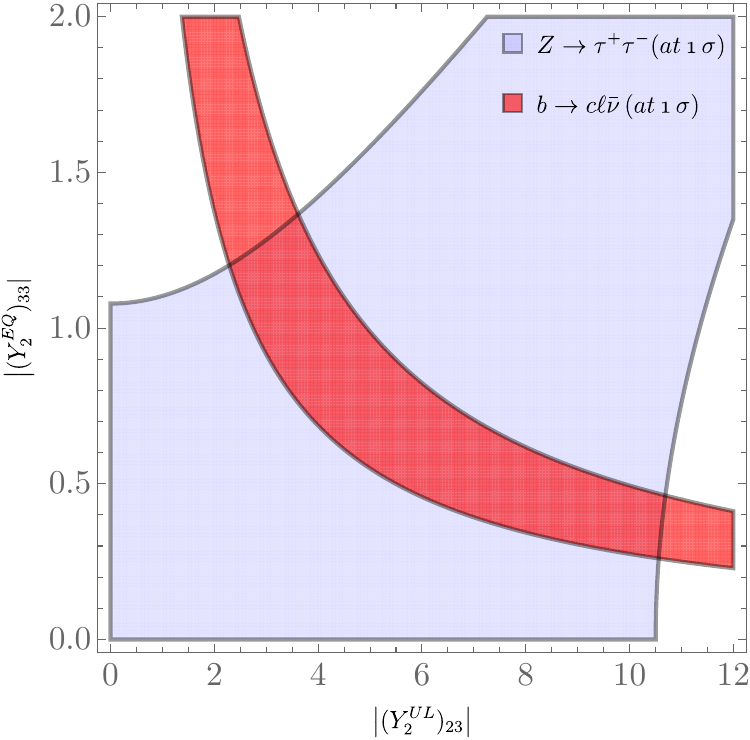}
\caption{Allowed region in the $|(Y^{UL}_2)_{23}|$ -- $|(Y^{EQ}_2)_{33}|$ plane based on constraints from $\mathcal{B}(Z\to
\tau^+ \tau^-)$ (gray, light shaded) and $b\to c\ell^-{\bar\nu}$ (red, dark shaded) for $m_{R_2}= 2 \, \text{TeV}$. \label{Bsab}}
\label{fig:ztautau}
\end{figure}
The LQ loop contributions are described by the
effective coefficients $\delta g_{f_{L(R)}}^{\tau \tau}$ that are
given as
\begin{align}
\label{eq:gL-F0-final}
\begin{split}
\Big{[}\delta g_{\tau_{L}}^{\tau\tau} \Big{]}_{F=0} \simeq & \,
N_C  \dfrac{|(Y_{2}^{UL})_{23}|^2}{16 \pi^2} \Bigg{[} \frac{1}{2}
\frac{x_{t}\left(x_{t}-1-\log x_t
\right)}{(x_{t}-1)^{2}}+\frac{x_{Z}}{12}F_{F=0}^{L}(x_{t})\Bigg{]}\,,
\end{split}
\end{align}
\begin{align}
\label{eq:gR-F0-final}
\begin{split}
\Big{[}\delta g_{\tau_{R}}^{\tau\tau} \Big{]}_{F=0} \simeq N_C  \dfrac{|(Y_{2}^{EQ})_{33}|^2}{16 \pi^2} \Bigg{[} -\frac{1}{2}
\frac{x_{t}\left(x_{t}-1-\log x_t
\right)}{(x_{t}-1)^{2}}+\frac{x_{Z}}{12}F_{F=0}^{R}(x_{t})\Bigg{]}\\+ x_Z\, N_C
\dfrac{|(Y_{2}^{EQ})_{3 3}|^2}{48 \pi^2} \Bigg{[} (-\frac{1}{2}+\frac{1}{3}\sin^{2}{\theta_{W}})
\left(\log x_Z-i\pi
-\dfrac{1}{6}\right)+\dfrac{(\frac{1}{2}-\frac{2}{3}\sin^{2}{\theta_{W}})}{6}\Bigg{]}\,
\end{split}
\end{align}
where $x_Z=m_Z^2/m_{R_2}^2$, $N_C=3$, with
$g_L^f=I_3^f-Q_f \sin^2 \theta_W$, $g_R^f = -Q_f \sin^2 \theta_W$
($f=u,d,\ell$) and $Q=Y+I_3$
while assuming $(Y_{2}^{EQ})_{33}$ and $(Y_{2}^{UL})_{23}$ are the only non-zero elements in our case.

In our analysis, the relevant combination of Yukawa couplings that explains the charged current $B$ anomalies is identified as $(Y^{UL}_2)_{23}(Y^{EQ}_{2})_{33}$. We constrain this product by setting all other Yukawa couplings to zero. In Figure~\ref{Bsab}, the allowed region for $|(Y^{UL}_{2})_{23}|$–$|(Y^{EQ}_{2})_{33}|$ plane, shown in gray, is determined from the constraint imposed by the branching fraction $\mathcal{B}(Z \to \tau \bar{\tau})$~\cite{ParticleDataGroup:2024cfk}, assuming $m_{R_2} = 2~\mathrm{TeV}$. We see that the allowed region that explains the charged-current $B$ anomalies (red, dark shaded region) gets constrained by $Z\to \tau^{+}\tau^{-}$-decay width measurements (gray, light-shaded region), disallowing $\vert(Y_{2}^{UL})_{23}\vert\gsim 10 $ and $\vert(Y_{2}^{EQ})_{33}\vert\gsim 1 $ in Fig~\eqref{fig:ztautau}.

%%%%%%%%%%%%%%%%%%%%%%%%%%%%%%%%%%%%%%%%%%%%%%%%%%
\section{Gauge Coupling Unification in Non-Minimal $SU(5)$}
\label{app:unif}

In this appendix, we briefly illustrate how the additional scalar degrees of freedom present in non-minimal SU(5) with a $45$-dimensional Higgs representation affect the running of the SM gauge couplings. Our aim here is purely illustrative: we are not claiming that gauge coupling unification can be naturally achieved in this non-supersymmetric setup with TeV-scale leptoquarks and diquarks. As emphasized below, a successful unification pattern in this framework typically requires highly tuned scalar mass spectra and leads to a relatively low unification scale, in tension with proton decay limits \cite{Super-Kamiokande:2020wjk}.

In addition to the SM gauge bosons and fermions, we consider a spectrum that contains a second Higgs doublet (as in a two-Higgs doublet model or 2HDM), a leptoquark $R_2$, a diquark $S_6$, as well as scalar multiplets originating from the $24_H$ and $45_H$, such as an $SU(2)_L$ triplet $T_3$ and a color octet $S_8$. Such states appear naturally in effective non-supersymmetric realizations of SU(5) with a $45$-representation scalar. By tuning the mass scales of these particles, we demonstrate a way to achieve gauge coupling unification.

At the one-loop level, the renormalization group equations (RGEs) for the gauge couplings are given by
\beq
\frac{d g_i}{d t} = \frac{b_i}{16 \pi^2}\,g_i^3\,, \qquad
t = \ln \left( \frac{\mu}{\mu_0} \right)\,,
\eeq
where \(b_i\) are the one-loop beta-function coefficients for the gauge group \(G_i\), \(g_i\) are the corresponding running gauge couplings, and $t$ is the logarithmic renormalization scale. Here $i = 1,2,3$ denote the SM gauge groups $U(1)_Y$, $SU(2)_L$, and $SU(3)_c$, respectively. The coefficients \(b_i\) depend on the gauge structure and the particle content of the theory. For a generic gauge group $G_i$ the coefficients \(b_i\) can be written as
\beq
b_i = -\frac{11}{3} C_2(G_i)
      + \frac{4}{3} \sum_{\text{fermions}} T(R_f)
      + \frac{1}{3} \sum_{\text{scalars}} T(R_s)\,,
\label{eq:general_b}
\eeq
where $C_2(G_i)$ is the quadratic Casimir invariant of the adjoint representation of $G_i$, and $T(R_f)$, $T(R_s)$ denote the Dynkin indices of the fermion and scalar representations $R_f$ and $R_s$, respectively.

To illustrate the impact of the extended scalar sector on the gauge coupling evolution, we divide the running into several energy intervals. We start from the electroweak scale and introduce new degrees of freedom at the respective thresholds of heavier particles. As a reference, the one-loop coefficients in the pure SM at scales just above the electroweak scale are \cite{Khalil:2014gba,FileviezPerez:2008afb},
\beq
b^{\rm SM} = (b_1, b_2, b_3)
           = \left(\frac{41}{10},\, -\frac{19}{6},\, -7\right)\,.
\eeq
We include a color-octet and iso-doublet scalar, $S_8$, that is a part of the $45_H$ at $\sim$ 400 GeV. The beta-function coefficients of the $S_8$ are (see Table I of Ref.~\cite{FileviezPerez:2008afb}),
\beq
b^{S_8}
  = \left(\frac{4}{5},\,\frac{4}{3},\,2\right)\,.
\eeq
 
At roughly the TeV scale, we include threshold corrections from a second Higgs doublet, the leptoquark $R_2$, the diquark $S_6$, and a color- and iso-triplet scalar, $S_3$, that also appears in the $45_H$. The cumulative effect of including all these particles results in the following beta function coefficient above the TeV scale,
\bea
b^{\rm 2HDM}
  &=& \left(\frac{1}{10},\,\frac{1}{6},\,0\right)\,, \\
b^{R_2} &=& \left(\frac{49}{30},\,\frac{1}{2},\,\frac{1}{3}\right)\,, \\
b^{S_6} &=& \left(\frac{2}{15},\,0,\,\frac{5}{6}\right)\,, \\
b^{S_3} &=& \left(\frac{1}{5},\,2,\,\frac{1}{2}\right)\,, \\
b^{{\rm SM} + S_8 + {\rm 2HDM} + R_2 + S_6 + S_3} &=& \left(\frac{209}{30},\,\frac{5}{6},\,-\frac{10}{3}\right)\,.
\eea
Finally, at the high scale of $\sim 4\times 10^7$ GeV, we add a color-singlet iso-triplet scalar, $T_3$, which appears in the $24_H$ representation of the non-minimal SU(5). This particle only has a non-zero beta-function coefficient for the $SU(2)_L$ gauge coupling, $b_2 = 1/3$.
\begin{figure}[h]
\centering
\includegraphics[width=0.7\textwidth]{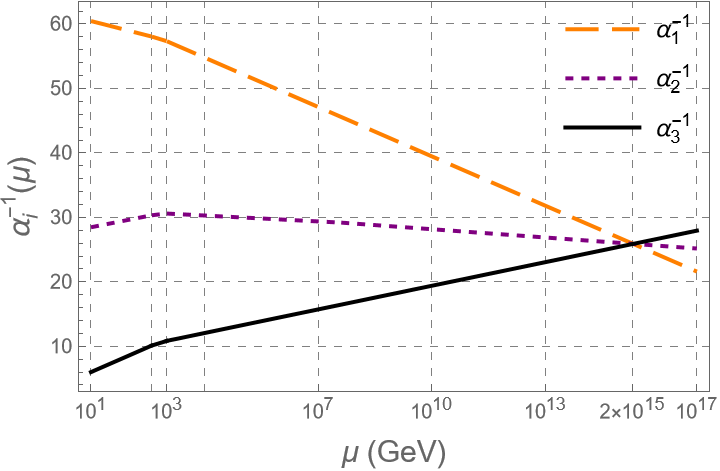}
\caption{Illustrative one-loop running of the gauge couplings in the non-supersymmetric
SU(5) model with a 45-dimensional Higgs representation and TeV-scale $R_2$ and $S_6$.
We plot $\alpha_i^{-1}(\mu) = 4\pi/g_i^2(\mu)$ as a function of the renormalization scale
$\mu$. The orange (dashed) line corresponds to the $U(1)_Y$ coupling, the purple (dotted) line
to the $SU(2)_L$ coupling, and the black (solid) line to the $SU(3)_c$ coupling.
Although the three couplings can be made to approach one another at a scale of order
$10^{15}$--$10^{16}$~GeV for a suitably tuned scalar spectrum, the resulting unification
scale is relatively low and cannot be considered a robust prediction of this non-supersymmetric
scenario. \label{fig:unif}}
\end{figure}
Using these coefficients in the appropriate energy intervals, we compute the one-loop 
evolution of the inverse gauge couplings, $1/\alpha_i(\mu) = 4\pi/g_i^2(\mu)$,
where the SU(5)-normalized hypercharge coupling is employed. In this convention, 
the \(U(1)_Y\) gauge coupling is rescaled according to $g_1^{\,2} = \frac{5}{3}\, g_Y^{\,2}$.
A representative example of the running in this non-supersymmetric SU(5) framework with the scales discussed above (i.e., $M_{S_8} \sim 400$ GeV, $M_{H_2} \sim M_{S_6} \sim M_{R_2}\sim M_{S_3} \sim 1$ TeV, and $M_{T_3} \sim 4 \times 10^7$ GeV), is shown in Fig.~\ref{fig:unif}. 

In this tuned spectrum, the three gauge couplings can be brought relatively
close to one another at a scale of order
\beq
M_{\rm GUT} \simeq 2 \times 10^{15}~\text{GeV}\,,
\qquad
\alpha_{\rm GUT}^{-1} \sim \mathcal{O}(30)\,.
\eeq
However, the unification scale remains comparatively low and is difficult to reconcile with current proton decay limits. Moreover, achieving such a pattern requires delicate adjustments of scalar
masses over widely separated scales, including light color octets and additional $SU(2)_L$ triplets, which undermines the naturalness and predictivity of the model. For these reasons, we do not rely on gauge coupling unification as a robust prediction of this non-supersymmetric non-minimal SU(5) setup; the discussion in this appendix should instead
be viewed as an illustration of how the extended scalar sector impacts the gauge coupling running.

\bibliography{su5bib}
\bibliographystyle{apsrev4-2-titles}

\end{document}